\documentclass[10pt,journal]{IEEEtran}
\usepackage{balance}
\usepackage{amssymb}
\usepackage{stmaryrd}
\usepackage{cite}
\usepackage{color}
\usepackage{multirow}
\usepackage{graphicx,times}
\usepackage{epstopdf}
\usepackage{indentfirst}
\usepackage{CJK}
\usepackage{amsmath}
\usepackage{amsfonts}
\usepackage{txfonts}
\usepackage{mathrsfs}
\usepackage{subfigure}
\usepackage{graphicx}
\usepackage{theorem}
\usepackage{algorithm}
\usepackage{algorithmic}
\usepackage{url}

\begin{document}
	
\title{Detecting Colluding Sybil Attackers in Robotic Networks using Backscatters}
\author{\IEEEauthorblockN{Yong Huang, Wei Wang,~\IEEEmembership{Senior Member,~IEEE}, Tao Jiang,~\IEEEmembership{Fellow,~IEEE}, Qian Zhang,~\IEEEmembership{Fellow,~IEEE}}
	\thanks{Part of this work has been presented at IEEE INFOCOM 2020 \cite{yong2020sybil}.}
	\thanks{This work was supported in part by the National Key R\&D Program of China under Grant 2020YFB1806606, National Science Foundation of China with Grant 62071194, 61729101, 91738202, Young Elite Scientists Sponsorship Program by CAST under Grant 2018QNRC001.\textit{ (Corresponding author: Wei Wang.)  }}
	\thanks{Y. Huang, W. Wang and T. Jiang are with the School of Electronic Information and Communications, Huazhong University of Science and Technology, Wuhan 430074, China (e-mail:\{yonghuang, weiwangw, taojiang\}@hust.edu.cn).}
	\thanks{Q. Zhang is with Department of Computer Science and Engineering, Hong Kong University of Science and Technology, Clear Water Bay, Hong Kong, China (e-mail:qianzh@cse.ust.hk).}}

\maketitle

\begin{abstract}
Due to the openness of wireless medium, robotic networks that consist of many miniaturized robots are susceptible to Sybil attackers, who can fabricate myriads of fictitious robots. Such detrimental attacks can overturn the fundamental trust assumption in robotic collaboration and thus impede widespread deployments of robotic networks in many collaborative tasks. Existing solutions rely on bulky multi-antenna systems to passively obtain fine-grained physical layer signatures, making them unaffordable to miniaturized robots. To overcome this limitation, we present ScatterID, a lightweight system that attaches featherlight and batteryless backscatter tags to single-antenna robots for Sybil attack mitigation. Instead of passively ``observing'' signatures, ScatterID actively ``manipulates'' multipath propagation by exploiting backscatter tags to intentionally create rich multipath signatures obtainable to single-antenna robots. Particularly, these signatures are used to carefully construct similarity vectors to thwart advanced Sybil attackers, who further trigger power-scaling and colluding attacks to generate dissimilar signatures. Then, a customized random forest model is developed to accurately infer the identity legitimacy of each robot. We implement ScatterID on the iRobot Create platform and evaluate it under various Sybil attacks in real-world environments. The experimental results show that ScatterID achieves a high AUROC of 0.987 and obtains an overall accuracy of 95.4\% under basic and advanced Sybil attacks. Specifically, it can successfully detect 96.1\% of fake robots while mistakenly rejecting just 5.7\% of legitimate ones.
\end{abstract}


\section{Introduction}

The continuous advancement of wireless technologies has promised to facilitate effective collaboration among a team of small and agile robots, which enables a wide spectrum of compelling applications, such as surveillance~\cite{rybski2002performance}, consensus~\cite{jadbabaie2003coordination}, aerial wireless coverage~\cite{chowdhery2018aerial} and search and rescue~\cite{zhang2006novel}. Although the openness of wireless medium delivers on the promise for efficient and agile collaboration, it also exposes robots to cyber attacks. A particular detrimental attack in robotic networks is the Sybil attack, which easily subverts the fundamental trust assumption in robotic collaboration by forging a large number of fake identities to gain a disproportionate influence in the network~\cite{newsome2004sybil,wang2018ghost}. For instance, by forging many IDs with excessive demands, a Sybil attacker can easily deplete valuable bandwidth resources from other genuine agents in a robotic network.

However, due to the ad hoc, dynamic, and miniaturized characteristics of robotic platforms, the Sybil attack mitigation in robotic networks still remains to be a challenging issue. Traditional pre-shared key (PSK) management schemes presume prior trust among network nodes~\cite{ramkumar2005an,wu2013fast,wang2018resonance,wang2018securing}, which are difficult to implement in ad hoc robotic networks where robots often go in and out. Alternatively, research efforts~\cite{demirbas2006an,xiao2009channel-based,liu2015the,faria2006detecting,xiong2013securearray,gil2017guaranteeing} measure received signals strength indicator (RSSI), channel state information (CSI) and angle of arrival (AoA) features in wireless physical layer (PHY) to verify the spatial uniqueness of each node. However, RSSI and CSI based techniques~\cite{demirbas2006an,xiao2009channel-based,liu2015the,faria2006detecting} not only need collaboration among multiple receivers or antennas, but also require all nodes to be stationary or semi-stationary. Although fine-grained AoA signatures can be extracted to detect nodes in close proximity under dynamic channels~\cite{xiong2013securearray,gil2017guaranteeing}, these approaches either rely on large multi-antenna arrays~\cite{xiong2013securearray} or require unnecessary robotic motions, such as in-place spin with two antennas~\cite{gil2017guaranteeing}. They are ill-suited to a team of robots individually with limited payload and hardware capabilities. 

In this paper, we argue that the fundamental hurdle in realizing lightweight Sybil-resilient solutions lies in that these PHY-based innovations focus on passively ``observing'' signal propagation signatures, which require bulky multi-antenna systems to capture fine-grained information. This paper explores a new approach: can we instead actively ``manipulate'' multipath propagation, to make conventionally multi-antenna-exclusive signatures also obtainable to a single antenna? If we could alter multipath propagation by just attaching several featherlight and batteryless backscatter tags to existing single-antenna robots, it would not require any hardware modification or incur load burden to the robots. 

\begin{figure}
	\centering
	\includegraphics[width=0.95\linewidth]{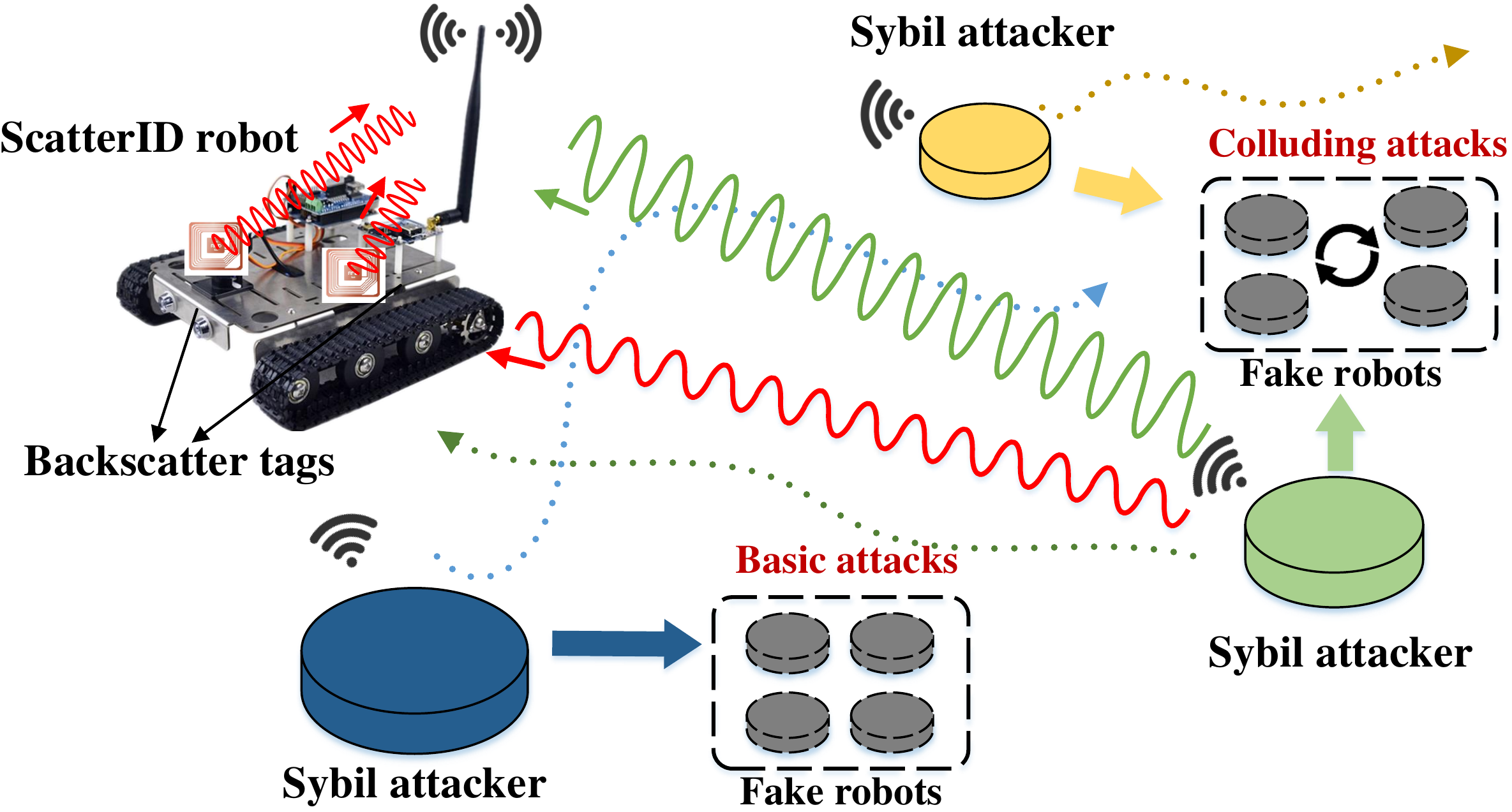}
	\caption{Illustration of Sybil attacks in a robotic network. Robotic attackers can launch various types of Sybil attacks.}
	\label{fig:observation}
\end{figure}

This paper proposes a lightweight Sybil attack detection system, ScatterID, which attaches featherlight and batteryless backscatter tags to single-antenna robots for defeating Sybil attackers who are even capable of launching advanced power-scaling and colluding attacks. Our fundamental insight is that when backscatter tags communicate by intermittently absorbing and reflecting ambient radio signals, the multipath between a pair of transceivers changes correspondingly~\cite{liu2013ambient}. Such fast changes, i.e., reflections from tags, provide unique spatial properties of the communication pair. In particular, backscattered signal strengths are highly correlated to distances between transceivers with respect to tags. By affixing several tags on a robot, as shown in Fig.~\ref{fig:observation}, other robots' trajectory information can be conveyed in backscattered signal traces. The spatial correlation indicates that similar backscatter signatures come from the same moving robot with a high probability. 
In this way, a ScatterID robot can exploit backscatter signal traces to generate unforgeable PHY IDs for its neighboring robots and thereafter discover fictitious ones among them by finding similar PHY IDs.

We realize the above idea by tackling the following two challenges.

\textit{1) How to exploit backscatter tags to construct sensitive signal profiles for hardware-constrained and dynamic robots?} With merely one antenna, a robotic platform cannot acquire fine-grained signal signatures, such as AoA information, for reliable attack detection in dynamic channel states. To deal with this challenge, we take sequential backscatter signatures of each robotic transmitter as a signal profile to characterize its long-term spatial information. Specifically, when hearing data transmission from a surrounding robot, a ScatterID robot controls its tags to reflect wireless signals in turn for avoiding overlapping among their backscattered signals. These backscattered signals are unique to the transmitter's location. In this way, the ScatterID robot continuously reflects a series of transmissions to obtain a signal profile that is sensitive to the transmitter's trajectory. 

\textit{2) How to effectively detect the presence of fake robots based on backscatter signatures under various Sybil attacks?} A Sybil attacker can scale its transmission power for each fake ID, making its backscatter signature different from others' in each transmission. In addition, attackers can collude with each other and randomly switch fake IDs in each transmission to disturb their long-term spatial similarity. Consequently, these attacks will incur significantly different signal profiles. To defend against such detrimental attacks, we propose an effective algorithm for attack-resilient robot verification. In particular, it first leverages Cosine distance to measure angular differences between any two profiles while mitigating power-scaling attacks. Then, it picks out representative features to reliably indicate each robot's similarity with respect to others for detecting colluding Sybil attackers. Based on extracted similarity vectors, a customized random forest model is designed to verify the legitimacy of each robot.

\textbf{Summary of Results.} We implement our system with iRobot Create robots, GNURadio/USRP B210 and backscatter tags using commercial off-the-shelf circuit components, and extensively evaluate our system under various Sybil attacks in typical indoor and outdoor environments. The evaluation results show that our system obtains an accuracy of 95.5\% under basic and power-scaling attacks and an accuracy of 95.3\% under colluding attacks. Overall, it can successfully detect 96.1\% of fake robots and meanwhile mistakenly reject just 5.7\% of legitimate robots.

\textbf{Contributions.} The main contributions of this work are summarized as follows. 
\begin{itemize}
	\item {We propose ScatterID, a lightweight Sybil attack detection system. It exploits featherlight and batteryless backscatter tags to actively create rich multipath signatures, which are obtainable to single-antenna robots.}
    \item {ScatterID effectively measures angular distances among signatures of different robots and further selects representative features from them, which makes it resilient to Sybil attackers with power-scaling and colluding abilities.}
    \item {ScatterID develops a customized random forest model that suits for backscatter signatures, which enables our system to accurately detect the existence of Sybil attackers while maintaining low computational overhead.}
    \item {We implement ScatterID on commercial robotic platforms and evaluate it under various Sybil attacks. The results show that ScatterID has effective resistance to not only basic and power-scaling attacks but also colluding attacks that are even never shown in the training phase.}
\end{itemize}

\section{Backscatter Enabled Signatures for Sybil Attack Detection}\label{sec:motivation}

\subsection{Sybil Attacks in Robotic Networks}
We focus on a general multi-robot network, where a team of miniaturized mobile robots coordinates their actions by exchanging information with each other in an ad hoc manner~\cite{jadbabaie2003coordination,cunningham2012fully}. In the robotic team, each robot has limited payload and hardware capabilities, and it is equipped with only one antenna for data transmission and reception. Moreover, all mobile robots have distinctive moving paths within a certain area to perform a given task, since multi-robot systems are usually spatially distributed to complete their tasks in a time-efficient way~\cite{yan2013survey}.

In the robotic network, we consider that one robot communicates with its $ N $ neighboring robots during task performing. However, some of these robots may be fictitious and originate from the same robot entity. Such attacks can be launched by Sybil attackers to gain unfair influence in the robotic network at low cost~\cite{newsome2004sybil}. Due to the ad hoc and dynamic characteristics of robotic platforms, traditional PSK schemes are difficult to maintain and therefore PHY signatures are widely exploited to defend against Sybil attacks in wireless networks. Hence, this work considers not only the basic Sybil attack but also two advanced Sybil attacks, which are launched by smart adversaries to increase the chance of avoiding PHY-based detection. The advanced Sybil attacks in the robotic network include power-scaling attacks and colluding attacks~\cite{xiao2009channel-based,liu2015the}. Specifically, let us assume that $ N_c $ Sybil attackers fabricate $ N_f $ fake robots in $ N $ neighboring robots. Three types of Sybil attacks are defined as follows.
\begin{itemize}
			\item \textbf{Basic Attacks.} Each attacker sends messages using multiple fake IDs for masquerading many fictitious robots and obtaining disproportionate resources in the network. 
			\item \textbf{Power-Scaling Attacks.} A Sybil attacker varies its transmit power for each fake ID in every transmission. Such attacks can accordingly change the power of received signals and render fake IDs to have backscattered signals with significantly different amplitudes.
			\item \textbf{Colluding Attacks.} $ N_c $ Sybil attackers collude with each other and share $ N_f $ fake IDs during task performing. Specifically, all attackers select $ N_c $ non-overlapping ID subsets from $ N_f $ fake IDs at each transmission slot and change their selections at different slots. Such attacks make fake IDs to change among different moving attackers and thus destroy the long-term similarity between PHY signatures of fake IDs.
\end{itemize}
The remaining $ N-N_f $ IDs that are not spawned are considered to be legitimate robots. 

\subsection{Characterizing Backscattered Signals} \label{sub:analysis}
Existing approaches to thwart Sybil attacks mainly rely on bulky multi-antenna systems, which are unaffordable for miniaturized robotic platforms. To overcome this dilemma, we advocate that lightweight and batteryless backscatter tags can provide fine-grained PHY signatures that are obtainable to single-antenna robots. According to~\cite{liu2013ambient}, when a transmitter is emitting a carrier signal $ \mathbf{x}(n) $, a receiver that is surrounded by $ K $ backscatter tags can receive the signals from the transmitter and tags at the same time. Thus, at the receiver side, the received signal $ \mathbf{y}(n) $ is a composite, which can be given as
\begin{align}\label{eq:backscatter channel}
	\mathbf{y}(n)=\mathbf{x}(n)+\sum_{k=1}^{K}\alpha_k\mathbf{b}_k(n)\mathbf{x}(n)+\mathbf{\eta}(n),
\end{align}
where $ \alpha_k $ and $\mathbf{b}_k(n) $ are, respectively, the reflection coefficient and the data bits of $k^{\text{th}}$ tag, and $ \mathbf{\eta}(n) $ is the environmental noise. In Eq.~\eqref{eq:backscatter channel}, $\mathbf{b}_k(n) $ is generated by switching the impedance of tag antenna between two states, which makes the $k^{\text{th}}$ tag to absorb and reflect incident signal $ \mathbf{x}(n) $ intermittently. 

The fast switching between tag states can dynamically change the multipath propagation between the transmitter and receiver, and such changes are highly correlated to the locations of transceivers with respect to tags. Since the multipath propagation between the transceiver pair is constant, we consider it as a compound path that is equivalent to a direct path in free space. Hence, we can focus on the multipath changes incurred by backscatter tags. Specifically, according to Friis path loss theory~\cite{rao2006theory}, the reflected signal strength $ P_{reflected} $ by the $k^{\text{th}}$ tag can be expressed as
\begin{align} \label{eq:friis path loss}
	P_{reflected}=\frac{P_t G_t}{4 \pi d_{kt}^{2}} \times \frac{\lambda^{2} G_r}{16 \pi^2 d_{kr}^{2}} \times T(\lambda,G_{k},\alpha_k),
\end{align}
where $ P_t $ is the transmission power, $ d_{kt} $ and $ d_{kr} $ the tag's equivalent distances to the transmitting and receiving antennas computed based on the compound direct path, $ G_t $ and $ G_r $ the transmitting and receiving antenna gains, respectively. Moreover, $ T(\cdot) $ is a function of the wavelength $ \lambda $, the antenna gain $ G_{k} $ of the tag and its reflection coefficient $ \alpha_k $. Based on Eq.~\eqref{eq:friis path loss}, we can find that the backscattered signal strength $ P_{reflected} $ is highly dependent on the transmission power and the relative distances of the $k^{\text{th}}$ tag in terms of transmitting and receiving antennas. By affixing the $k^{\text{th}}$ tag on a mobile robot, $ d_{kr} $ is a fixed distance, and $ P_t $ and $ d_{kt} $ are only variables in the Eq.~\eqref{eq:friis path loss}. Therefore, the reflected signal strength highly correlates with transmission power and the distance between the robotic communication pair. In this way, given a constant transmission power $ P_t $, we can construct a spatial-related signature relying on reflected signals from $ K $ backscatter tags.

\begin{figure}
	\centering
	\includegraphics[width=0.975\linewidth]{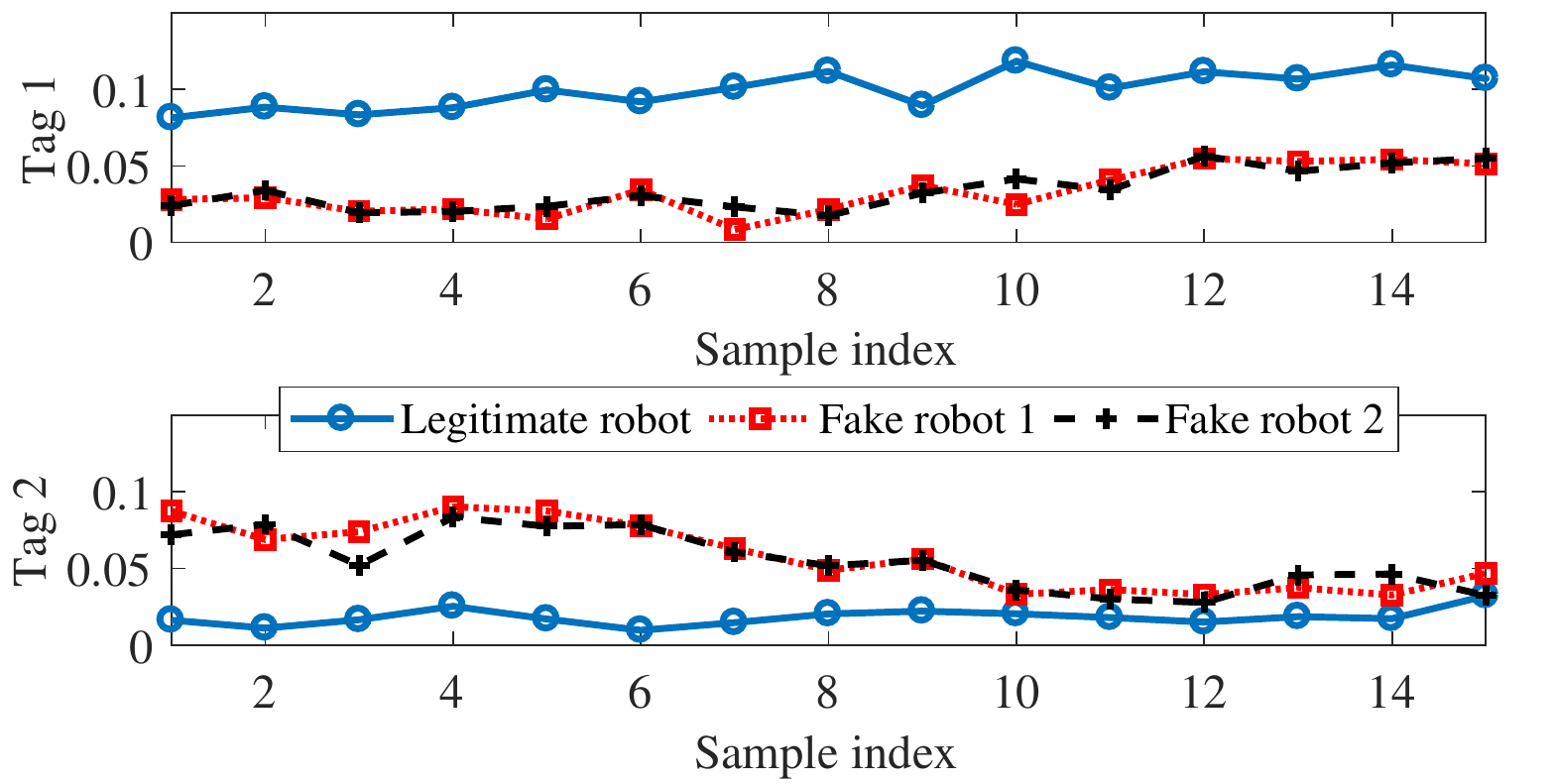}
	\caption{Backscattered signal traces reflected by two tags from one legitimate and two fake robots.}
	\label{fig:traces}
\end{figure}

\subsection{Feasibility Study}

We perform a set of preliminary experiments to verify the above idea using backscatter tags, moving robotic platforms and USRP nodes. To avoid interference between the carrier signal $ \mathbf{x}(n) $ and reflected signals $ \sum_{k=1}^{K}\alpha_k\mathbf{b}_k(n)\mathbf{x}(n) $ in Eq.~\eqref{eq:backscatter channel}, we implement frequency-shift backscatter tags based on recent work~\cite{kellogg2015wi-fi,zhang2016hitchhike,wang2017fm}. Moreover, the tags are set to reflect ambient signals in turns for avoiding overlapping among backscattered signals from different tags. In the experiment, we attach two tags to one USRP node who logs backscattered signal traces in a fixed position. We place two USRP nodes on different moving platforms to emulate one legitimate robot and one Sybil attacker with two different identities, respectively. Both the legitimate robot and Sybil attacker transmit packets with the same transmission power while moving around in different trajectories. We extract the backscattered signal traces of two tags and plot the corresponding results in Fig.~\ref{fig:traces}. 

From Fig.~\ref{fig:traces}, we can observe that for two fake robots, their backscattered signal traces are nearly identical to each other. This is because that from the same moving robot, their signals experience similar propagation paths that are created by two tags. Whereas, for a pair of legitimate and fake robots, their signal propagations from the transmitter to the backscatter tags are different due to distinctive trajectories. Thus, their backscattered signal traces are uncorrelated with each other. This observation verifies that the reflected signals from backscatter tags are highly sensitive to robotic trajectories, and they can be further utilized to construct a unique signal profile for each moving robot. 

\begin{figure}
	\centering
	\includegraphics[width=0.97\linewidth]{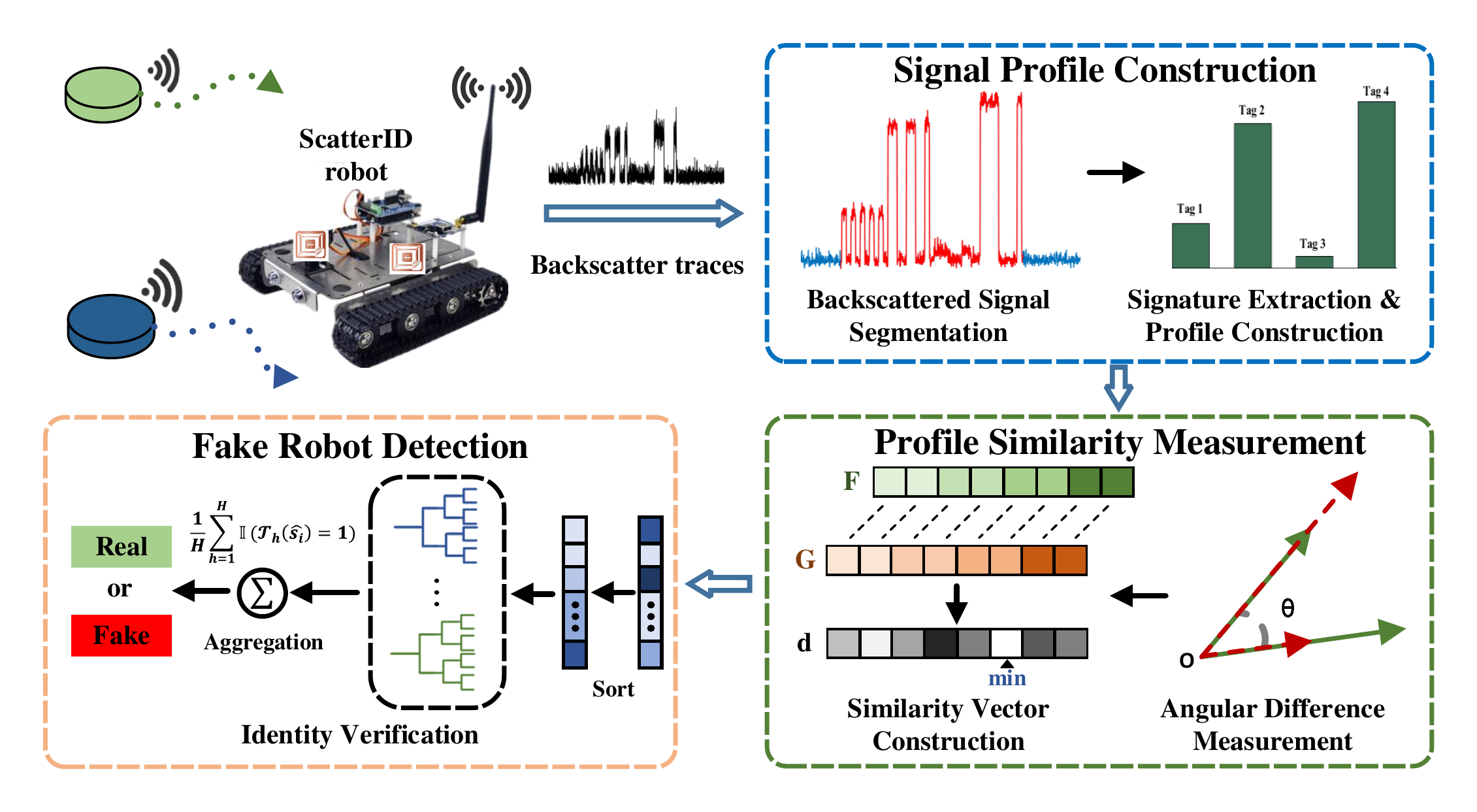}
	\caption{System flow of ScatterID. It contains three core components, i.e., Signal Profile Construction, Profile Similarity Measurement and Fake Robot Detection.} 
	\label{fig:system-overview}
\end{figure}

\section{Backscatter Signature Based Sybil Attack Detection}\label{sec:system design}

\subsection{System Overview} 
ScatterID is a lightweight system that provides effective resilience to various types of Sybil attacks in multi-robot networks. Specifically, when hearing data transmission from surrounding robots, a ScatterID robot controls its tags to backscatter their signals. Then, ScatterID extracts their signal profiles that capture robotic spatial information. Next, to mitigate potential power-scaling attacks, it calculates Cosine distances among signal profiles. Moreover, to deal with colluding attacks, it constructs a similarity vector for each profile by selecting the minimal value from distances with respect to the others in each transmission. Based on the similarity vector, ScatterID leverages a classification model to output a detection decision for each robot. If one surrounding robot is detected to be fictitious, the ScatterID robot terminates their connection link. Otherwise, the ScatterID robot continues their communication.

As illustrated in Fig.~\ref{fig:system-overview}, the core of our system includes three components -- \textit{Signal Profile Construction}, \textit{Profile Similarity Measurement} and \textit{Fake Robot Detection}.
\begin{itemize}
	\item  \textbf{Signal Profile Construction.} This component first segments the backscattered signal from the received signal, and it then effectively extracts the reflected signals from all tags as a multipath signature in each transmission. Then, a signal profile can be obtained as a sequence of signatures for each neighboring robot.
	\item  \textbf{Profile Similarity Measurement.} After profiling all neighboring robots, this component measures the angular differences among signal profiles using Cosine distance for defending against power-scaling attacks. Then, it extracts a vector of minimal distances from a distance matrix to effectively mitigate colluding Sybil attacks.
	\item  \textbf{Fake Robot Detection.} This component detects the presence of fake robots using a customized random forest model. In particular, it sorts each similarity vector in an ascent manner, inputs the sorted vector into multiple decision trees and aggregates all prediction results to make its final decision. 
\end{itemize}

\subsection{Signal Profile Construction}
The first step of ScatterID is to construct sensitive signal profiles for all surrounding robots. These profiles are carefully constructed based on a sequence of backscattered signals to reliably provide unique spatial properties of moving robots. 

\begin{figure}
	\centering
	\includegraphics[width=0.98\linewidth]{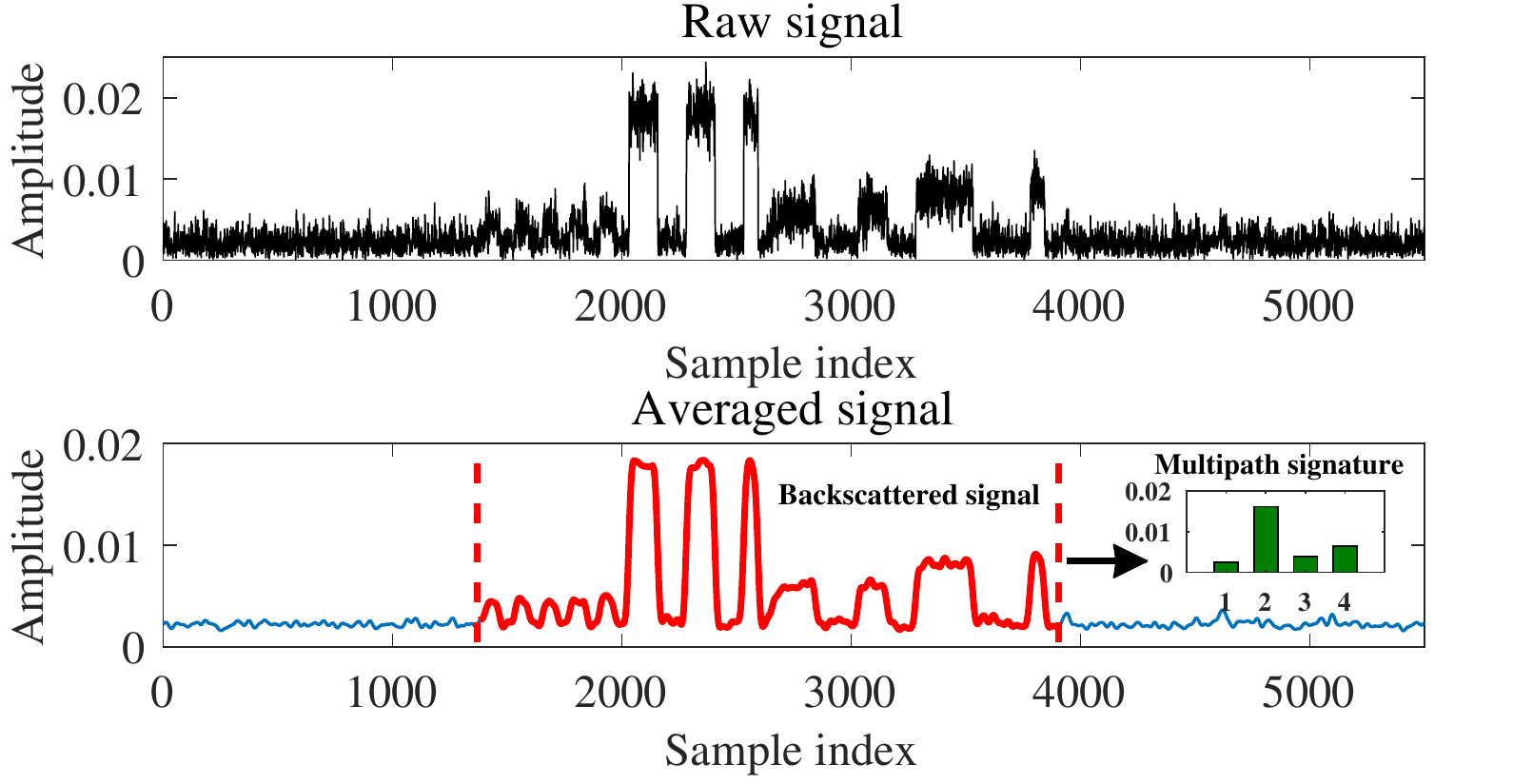}
	\caption{Backscattered signal segmentation and signature extraction. The red segment represents a backscattered signal, and the green bars indicate extracted backscatter signatures.}
	\label{fig:backscatteredsignalextraction}
\end{figure}

\textbf{Backscattered Signal Segmentation.} To update backscattered signals, a ScatterID robot controls all tags to reflect data packets in turns for avoiding overlapping between backscattered signals from different tags and thereafter extracts a segment of backscattered signal from received signals periodically. Since the received signal may contain the component without tags' reflections, it is highly desirable to detect and segment the backscattered component from the received signal. 

To achieve this goal, our system first adopts a moving average method~\cite{liu2013ambient} to smooth the received signal. Fig.~\ref{fig:backscatteredsignalextraction} plots an instance of a received signal and the corresponding smoothing result. As the figure shows, such smoothing operations enable our system to effectively remove signal noise and reliably decode transmitted message from tags. Next, we leverage tags' transmitted message to determine the start and end of backscattering in the averaged signal. Specifically, the transmitted message is a known binary array of zeros and ones, which are encoded in the backscattered signal to differentiate each tag's reflection from others. Thereby, the backscattered component of the averaged signal and the tags' binary array are highly correlated. Hence, we can correlate the averaged signal $ \mathbf{s}(n) $ with the binary array $ \mathbf{i}(n) $ to detect when tags begin reflecting signals. The correlation result $ \mathbf{c}(n) $ can be written as 
\begin{align}\label{coorelation}
	\mathbf{c}(n)=\sum^T_{t=1} \mathbf{s}(n+t-1) \times \mathbf{i}(t),
\end{align}
where $ T $ is the length of $ \mathbf{i}(n) $. According to Eq.~\eqref{coorelation}, $ \mathbf{c}(n) $ will have the highest peak when the backscattered signal and $ \mathbf{i}(n) $ completely overlap. Thus, our system identifies the backscattering start $ t_{start} $ by finding $ n$ with the maximum value in $ \mathbf{c}(n) $. Moreover, as the length of the backscattered signal equals to $ T $, we can accordingly obtain the end of backscattering as $ t_{end} = t_{start} + T $. After extracting $ t_{start} $ and $ t_{end} $, our system can accurately segment the backscattered signal $ \mathbf{B} $:
\begin{align} \label{eq: backscatter signal}
	\mathbf{B}=(\mathbf{b}_{1},\mathbf{b}_{2},\cdots, \mathbf{b}_{K}),
\end{align}
where $ \mathbf{b}_{i} $ is the reflected component by the $i^{\text{th}}$ tag and $ K $ is the number of tags used in our system.

\textbf{Signature Extraction.} Now that we have picked the backscattered component out of the received raw signal, we take the next step to extract a multipath signature from it. As depicted in Fig.~\ref{fig:backscatteredsignalextraction}, the backscattered signal consists of not only the tags' reflections but also the unwanted reflections from ambient environments. To extract reliable multipath features from each tag, ScatterID leverages the fact that the difference between reflected and non-reflected samples is caused by the tags' reflections. Therefore, by subtracting two kinds of samples, we can effectively eliminate environmental reflections and thereafter obtain the reflections from all tags. Mathematically, with the backscattered signal of the $i^{\text{th}}$ tag $ \mathbf{b}_{i}(n) $ in Eq.~\eqref{eq: backscatter signal}, we assume that there are $ N_1 $ reflected and $ N_0 $ non-reflected samples. Hence, the reflection caused by the $i^{\text{th}}$ tag $ p_i $ can be computed by
\begin{align}\label{power extraction}
	p_i = \frac{1}{N_1} \sum^{N_1}_{n=1} \mathbf{b}_{i}(n|ref.)-\frac{1}{N_0} \sum^{N_0}_{n=1} \mathbf{b}_{i}(n|non-ref.).
\end{align}

As previously analyzed in Section~\ref{sec:motivation}, the tag's reflection $ p_i $ highly depends on the distance between the tag and signal resource. Thus, with multiple backscatter tags, our system can construct a multipath signature that provides location information of the signal resource. Formally, based on $ K $ tags' reflections, a multipath signature $\mathbf{f} $ can be expressed as
\begin{align} \label{eq: multipath signature}
	\mathbf{f}=(p_1, p_2, \cdots, p_K).
\end{align}

\textbf{Profile Construction.} To this end, ScatterID can construct a signal profile for each neighboring robot. Due to the mobility of robotic platforms, multipath signatures from the same Sybil attacker may have a poor similarity in each transmission. To deal with this issue, our system periodically extracts a multipath signature from each robot's backscattered signal traces, and it takes a sequence of successive multipath signatures as a signal profile for reliably capturing its trajectory information. For fake robots forged by the same moving attacker, their backscattered signal traces have the same trend and are nearly identical to each other, as depicted in Fig.~\ref{fig:traces}. Hence, by characterizing long-term spatial information, their signal profiles will be similar to each other. Formally, a signal profile $ \mathbf{F} $ of each moving robot can be expressed as 
\begin{align} \label{eq: definition of profile}
\mathbf{F}=[\mathbf{f}_1; \mathbf{f}_2; \cdots; \mathbf{f}_L],
\end{align}
where $ \mathbf{f}_l $ denotes the $l^{\text{th}}$ multipath signature and $ L $ is the number of signatures.

\subsection{Profile Similarity Measurement} 

\begin{figure}
	\centering
	\includegraphics[width=0.97\linewidth]{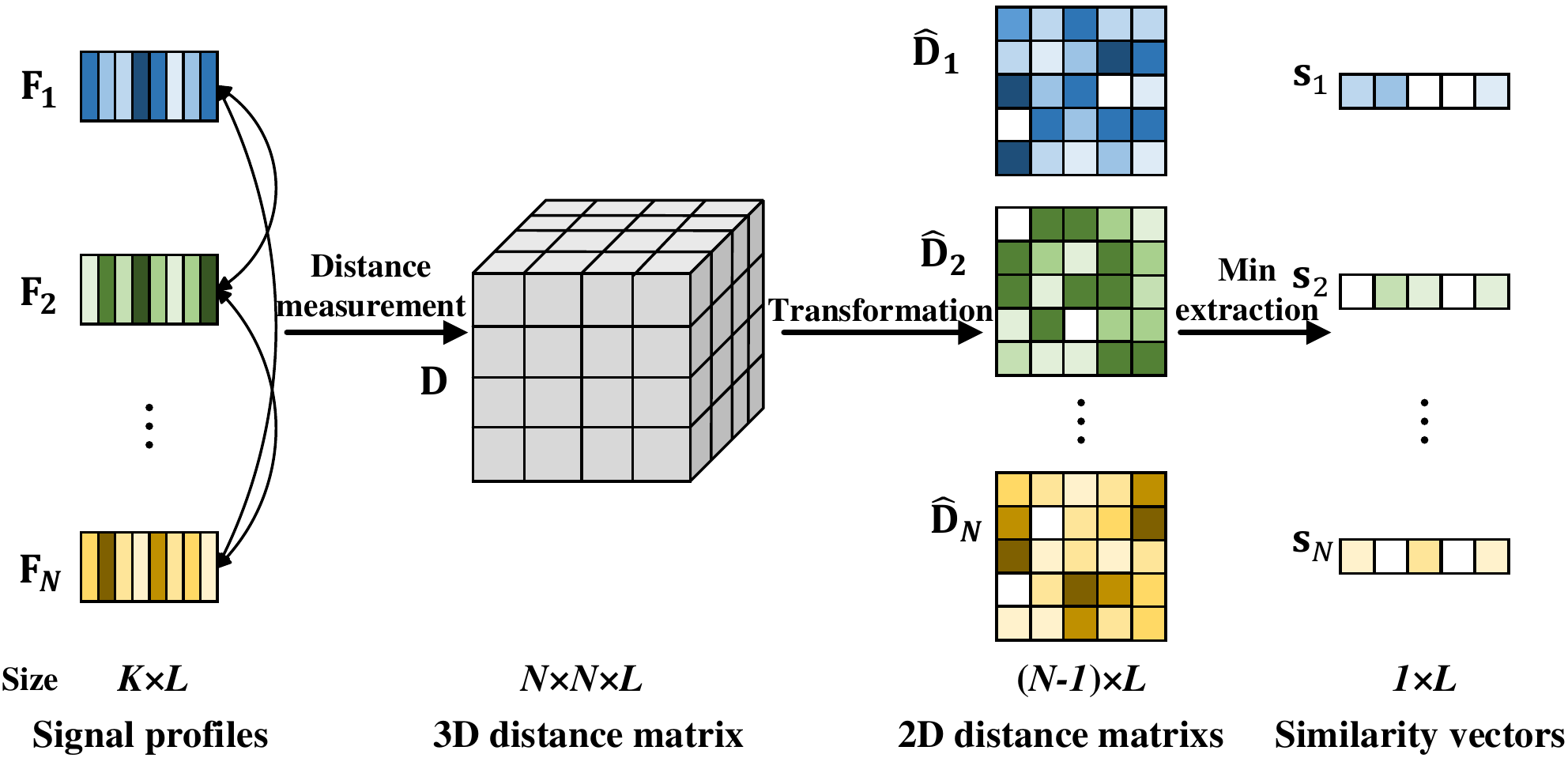}
	\caption{Work flow of the profile similarity measurement. $ N $ is the number of surrounding robots, $ K $ the number of tags and $ L $ the number of multipath signatures in a signal profile.}
	\label{fig:profile-similarity-measurement}
\end{figure}

In the robotic network, Sybil attackers can trigger power-scaling and colluding attacks. To mitigate these threats, ScatterID proceeds to transform their signal profiles into desirable similarity vectors that are resilient to various types of Sybil attacks as shown in Fig.~\ref{fig:profile-similarity-measurement}.

\textbf{Similarity Vector Construction.} Consider that there are $ N $ neighboring robots who communicate with the ScatterID robot, and a set of their signal profiles can be denoted as $\left\lbrace  \mathbf{F}_1, \mathbf{F}_2, \cdots, \mathbf{F}_N \right\rbrace $. Based on these profiles, we measure the distance between every two profiles and thus obtain a 3D distance matrix $ \mathbf{D} $ as
\begin{equation} \label{eq: distance matrix}
\mathbf{D}=\left[ \begin{array}{cccc}
0      & \mathbf{d}_{12}      & \cdots & \mathbf{d}_{1N}      \\
\mathbf{d}_{21}     & 0      & \cdots & \mathbf{d}_{2N}     \\
\vdots & \vdots & \ddots & \vdots \\
\mathbf{d}_{N1}      & \mathbf{d}_{N2}      & \cdots & 0      \\
\end{array} 
\right ],
\end{equation}
where $ \mathbf{d}_{mn} $ is the distance measurements with respect to two profiles $ \mathbf{F}_m $ and $ \mathbf{F}_n $. Since a signal profile is comprised of $ L $ backscatter signatures, ScatterID calculates the distances between $ \mathbf{F}_m $ and $ \mathbf{F}_n $ in a pairwise manner, and it thereafter outputs a $ L $-dimensional distance vector as 
\begin{align}\label{eq: distance vector}
	\mathbf{d}_{mn}=(d^{mn}_1,d^{mn}_2,\cdots, d^{mn}_L),
\end{align}   
where $ d^{mn}_l=\mathbf{dist}( \mathbf{f}_{ml},  \mathbf{f}_{nl}) $, the distance between the $l^{\text{th}}$ rows of $ \mathbf{F}_m $ and $ \mathbf{F}_n $. Compared to a scalar distance, the distance vector $ \mathbf{d}_{mn} $ contains the multipath signature differences between two moving robots within a certain period, and thus provides finer-grained information about their long-term similarity. To obtain all distances with respect to the  $n^{\text{th}}$ robot, we extract a 2D distance matrix from $ \mathbf{D} $ as
\begin{equation} \label{eq: distance matrix of each robot}
\mathbf{\hat{D}}_n=\left[ \begin{array}{c}
\mathbf{d}_{1n}\\
\mathbf{d}_{2n}\\
\vdots  \\
\mathbf{d}_{Nn} \\
\end{array} 
\right ]=\left[ \begin{array}{cccc}
d^{1n}_1      & d^{1n}_2      & \cdots & d^{1n}_L      \\
d^{2n}_1     & d^{2n}_2      & \cdots & d^{2n}_L     \\
\vdots & \vdots & \cdots & \vdots \\
d^{Nn}_1      & d^{Nn}_2      & \cdots & d^{Nn}_L      \\
\end{array} 
\right ],
\end{equation}
where $ \mathbf{\hat{D}}_n $ has a size of $ (N-1)\times L $.

For a group of smart attackers, they can collude with each other and launch Sybil attacks. By changing between different moving attackers, fake IDs will have more dissimilar signal traces in the transmission period, making themselves harder to be detected based on long-term signatures. To defend against such detrimental attacks, we construct a similarity vector for each robot. Specifically, we pick out the minimal distance in each column of $ \mathbf{\hat{D}}_n $ to indicate the similarity of the $ n^{\text{th}} $ robot with respect to the remaining robots in each transmission. Thus, we can obtain a similarity vector $ \mathbf{s}_n $ as
\begin{align}\label{eq: similarity vector}
\notag	\mathbf{s}_n & =\left(
	\mathop{\min } \left\lbrace d^{in}_{1}\right\rbrace_{1 \le i \le N} ,\mathop{\min } \left\lbrace d^{in}_{2}\right\rbrace_{1 \le i \le N}, \cdots, \mathop{\min } \left\lbrace d^{in}_{L}\right\rbrace_{1 \le i \le N}
	\right)\\ & =\left( 
	s^{n}_1, s^{n}_2,\cdots ,s^{n}_L
	\right),
\end{align}
where $ \left\lbrace d^{in}_{l}\right\rbrace_{1 \le i \le N} $ represents all elements of the $ l^{\text{th}} $ column in $ \mathbf{\hat{D}}_n $. The intuition behind this processing is that whether the triggered attacks are colluding or not, there always exist two fake IDs that come from the same attacker in each transmission, thus the distance between their signatures should be very small. In contrast, a legitimate ID has a different backscatter signature with the other IDs' in a high probability and consequently has a large signature distance from the others. Hence, via selecting the minimum within the distances in each transmission, a similarity vector $ \mathbf{s}_n $ would have large values for legitimate robots and small values for fake robots spoofed by Sybil attackers. 

\begin{figure}
	\centering
	\includegraphics[width=0.99\linewidth]{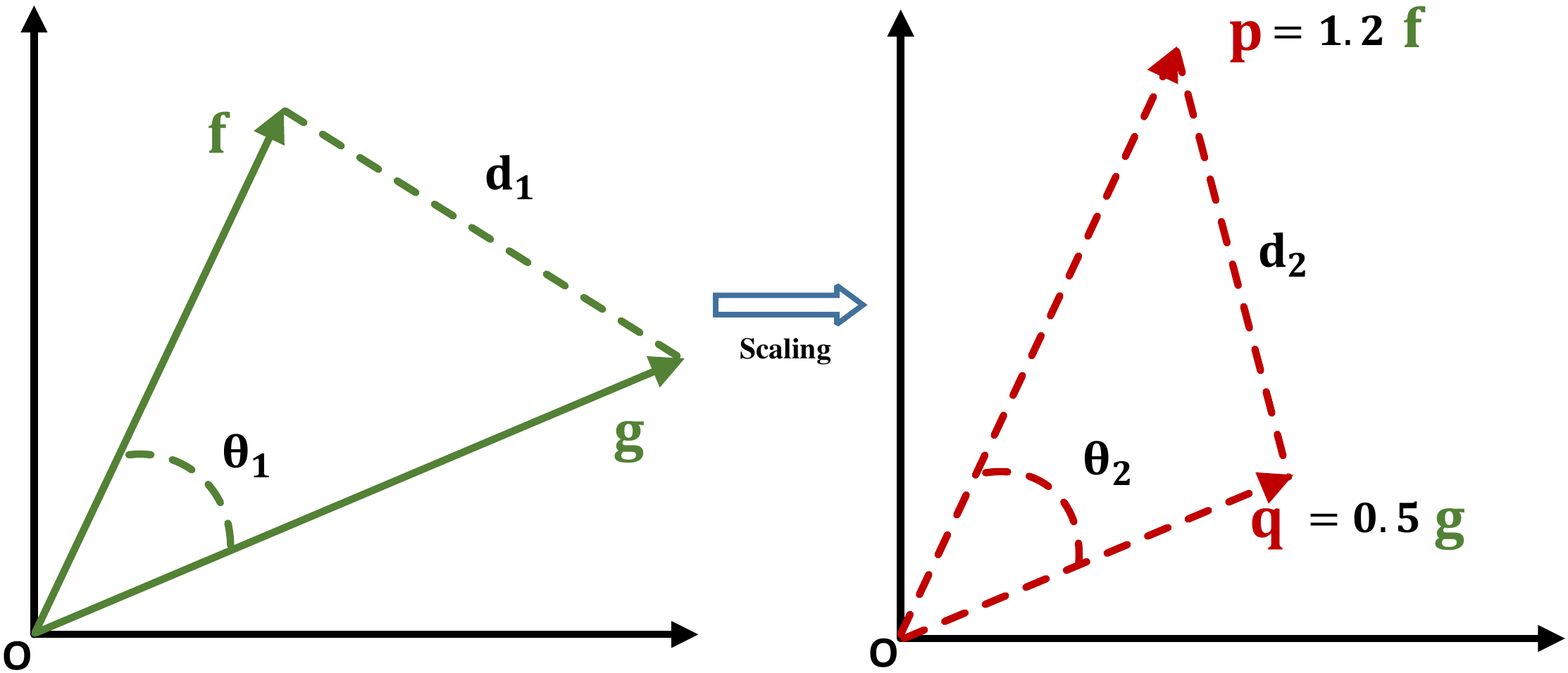}
	\caption{Illustration of power-scaling attacks and Cosine distance. After power-scaling attacks, Cosine distance remains the same, but Euclidean distance is distorted.}
	\label{fig:cosinedistance}
\end{figure}

\textbf{Angular Difference Measurement.} Now that we proceed to deal with power-scaling attacks, where a clever attacker changes its transmission power to emulate different fake robots. Specifically, according to Eq.~\eqref{eq:friis path loss}, by intentionally changing $ P_t $ in each transmission, the attacker can manipulate $ \mathbf{f} $ in Eq.~\eqref{eq: multipath signature} with a varying scale coefficient $ \alpha $ for each identity. Such an operation scales the reflected signals by all tags and further leads to a scaled signature $  \mathbf{f}' = (\alpha p_1, \alpha p_2, \cdots, \alpha p_K) $, which consequently renders our system ineffective in discovering two fake robots. Since a similarity vector $ \mathbf{s}_n $ consists of distances between two backscatter signatures and is the processing unit for subsequent identity verification, the distance metric $ \mathbf{dist}(\cdot, \cdot) $ should be carefully determined for resilience to power-scaling attacks.

As shown in Fig.~\ref{fig:cosinedistance}, after power-scaling attacks, two signature vectors $ \mathbf{f} $ and $ \mathbf{g} $ are both scaled into $ \mathbf{p} $ and $ \mathbf{q} $ in the feature space with different scale coefficients $ \alpha_1 = 1.2 $ and $ \alpha_2 = 0.5 $, respectively. Such transformation greatly changes the lengths of two signatures and distorts their Euclidean distance, i.e., $ d_1 \ne d_2 $, making Euclidean metric ineffective in capturing their similarity information. To find an effective metric, we observe that under power-scaling attacks, although the lengths of two signatures are distorted, their directions remain unchanged. Specifically, the directional difference of $ \mathbf{f} $ and $ \mathbf{g} $ are the same with that of $ \mathbf{p} $ and $ \mathbf{q} $, i.e., $ \theta_1 = \theta_2 $. Hence, instead of using the length difference, our system should rely on the directional difference between two backscatter signatures for distance measurement. For this purpose, we use Cosine distance to measure the directional difference between any two multipath signatures. Formally, as depicted in Fig.~\ref{fig:cosinedistance}, the Cosine distance between $ \mathbf{p} $ and $ \mathbf{q} $ is defined as
\begin{align}\label{cosine distance}
 \mathbf{dist}_{cos}(\mathbf{p},\mathbf{q} ) = 1- \frac{ \left< \mathbf{p}, \mathbf{q} \right>} { \left\|\mathbf{p}\right\|_2 \left\|\mathbf{q}\right\|_2 },
\end{align}
where $ \left< \cdot, \cdot \right> $ is the inner product and $ \left\| \cdot \right\|_2 $ the L2 norm. According to Eq.~\eqref{cosine distance}, the scaling coefficients $ \alpha_1 $ and $ \alpha_2 $ will be eliminated, making our system effective in mitigating power-scaling attacks from Sybil attackers.

\begin{figure}
	\centering
	\includegraphics[width=0.99\linewidth]{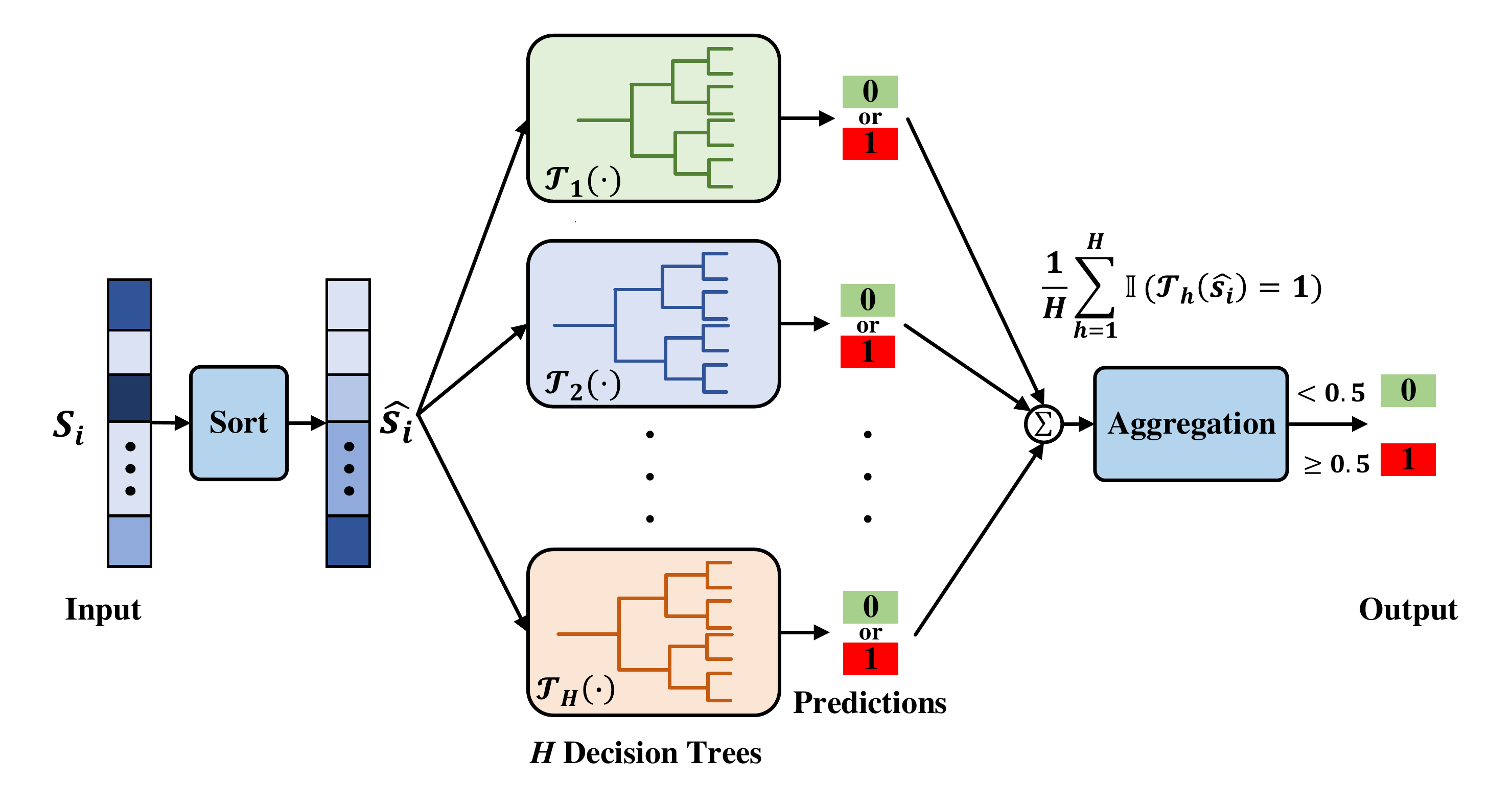}
	\caption{Our RF model for fake robot detection. "0" stands for a real robot and "1" for a fake one.}
	\label{fig:random-forest}
\end{figure}

\subsection{Fake Robot Detection}

Based on similarity vectors of all robots, we proceed to perform fake robot detection. 

\textbf{Identity Verification.} Given a similarity vector $ \mathbf{s}_{i} $, we would like to construct a detection function $ \mathcal{F} (\cdot) $ such that
\begin{align}\label{detection fucnction}
\mathcal{F}\left( \mathbf{s}_{i}\right) =\left\lbrace 
\begin{array}{lcl}
1 \: , & \text{if} & \text{$ i^{\text{th}} $ robot is fake,} \:  \\
0 \: , & \text{if} & \text{$ i^{\text{th}} $ robot is real.} \:
\end{array}
\right.
\end{align} 
For this purpose, we exploit a data-driven algorithm to fit $ \mathcal{F} (\cdot) $ with our labeled dataset $ \mathcal{C}= \left\lbrace (\mathbf{s}_{i},y_i) \right\rbrace_{i=1:M}  $, where $ M $ is the number of included samples and $ y_i \in \left\lbrace 0,1\right\rbrace  $ the corresponding label. In the component of fake robot detection, the selected detection algorithm should be lightweight to implement on robotic platforms with limited storage and computation capabilities. Besides, it should have high detection performance under high robotic movements and environmental dynamics.

To satisfy these requirements, we resort to the random forest (RF) model. Generally, the RF model is a well-known ensemble learning method used in many real-world applications, and it works by building a collection of biased decision trees and aggregating their results for a final outcome~\cite{Breiman2001Random}. On the one hand, for each classification decision, the RF model only needs to simply average outputs of all internal decision trees. Thus, its computational complexity is just linearly related to the number of lightweight and easily-implement decision tree models. On the other hand, instead of relying on one single base classifier, the RF model consults many relatively uncorrelated decision trees, which together act as a decision committee, and outputs the decision with the most votes. In this way, the final outcome of the RF model is more accurate and stable, and outliers caused by environmental noise has little impact on it. In our experiment, we compare the RF model with other widely-used machine learning models and show that it achieves the highest detection accuracy with a moderate testing time for each sample, which verifies its merits of high performance and low complexity. Hence, we adopt a RF model to perform the task of fake robot detection. Specifically, as shown in Fig.~\ref{fig:random-forest}, our RF model first makes a bunch of binary predictions from $ H $ decision trees with respect to $ \mathbf{s}_{i} $. Then, it aggregates all intermediate predictions and outputs the final outcome based on its decision criterion. 

\begin{figure}
	\centering
	\subfigure[Unsorted.]{
		\includegraphics[width=0.475\linewidth]{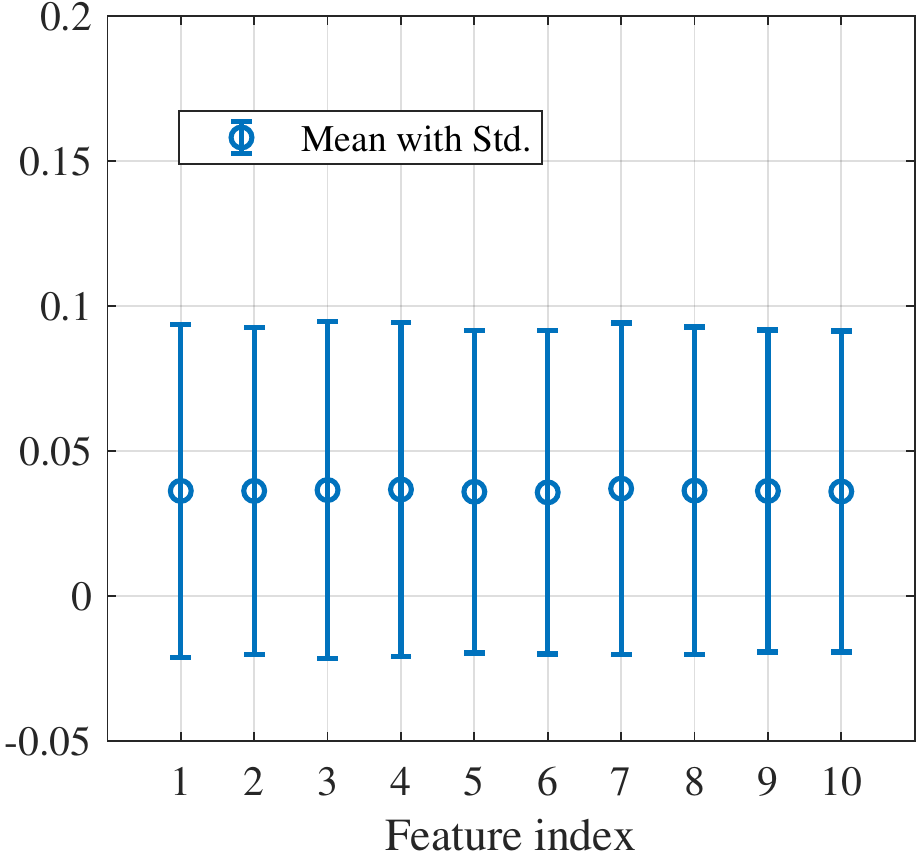}}
	\subfigure[Sorted.]{
		\includegraphics[width=0.475\linewidth]{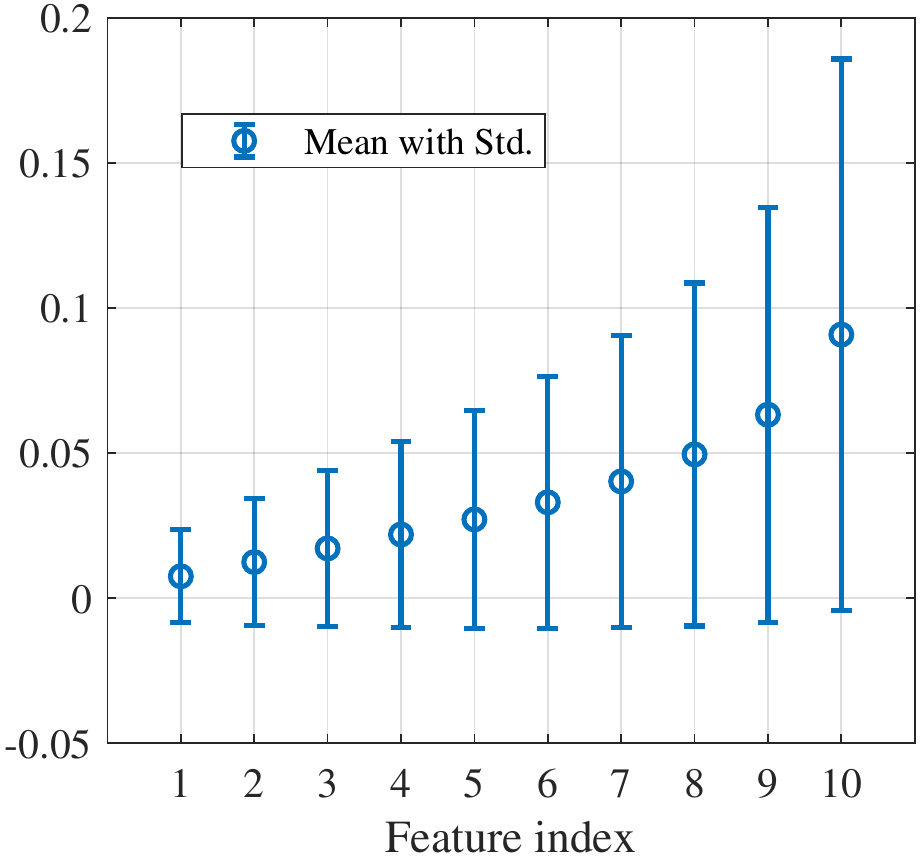}}
	\caption{Mean and standard deviation of sample features in our training dataset. The sort operation can effectively differentiate feature distributions.}
	\label{fig:Arttribute distributions } 
\end{figure} 

In the first step, we train $ H $ biased decision trees $ \left\lbrace \mathcal{T}_{1} (\cdot), \mathcal{T}_{2} (\cdot), \cdots,  \mathcal{T}_{H} (\cdot) \right\rbrace $ as base classifiers for label predictions. Since making decision trees more uncorrelated is essential for a RF model, two dimensions of randomness are introduced to decorrelate trees during tree building. First, we construct $ H $ independent data subsets $ \left\lbrace \mathcal{C}_1, \mathcal{C}_2, \cdots, \mathcal{C}_H  \right\rbrace  $ by sampling $ M $ times with replacement for each subset, known as bootstrapping, from the original dataset $ \mathcal{C} $ and train the tree $ \mathcal{T}_{h} (\cdot) $ on the subset $ \mathcal{C}_{h} $ only without pruning. The bootstrapping can introduce lower correlation between any two subsets and thus force their trees to capture different trends in the dataset $ \mathcal{C} $. Second, we randomly select $ 1 \le z \le L $ features out of a similarity vector when splitting tree nodes, i.e, making decision rules, to further introduce diversity between any two trees and randomness into the model. In our RF model, we set $ z=\log_2 L  $ according to~\cite{Breiman2001Random}. However, based on the definition in Eq.~\eqref{eq: similarity vector}, a similarity vector $ \mathbf{s}_{i} $ is consisted of a time series of distance features, and thus each feature, i.e., one attribute of samples, contains the same type of data points and would has a similar distribution with the others in the dataset $ \mathcal{C} $. To verify this, we compute the mean and standard deviation of each feature in our training dataset. As Fig.~\ref{fig:Arttribute distributions } (a) shows, all features have the nearly same statistics, implying that their distributions are very close to each other. This high similarity of features would enlarge the correlation between trees and eventually undermine the effectiveness of the random feature selection. To address this problem, we sort every $ \mathbf{s}_{i} $ in an ascent manner before tree building as 
\begin{align}
	\hat{\mathbf{s}_{i}} = \textbf{sort} (\mathbf{s}_{i}) = \left( \hat{s}^{i}_1, \hat{s}^{i}_2,\cdots ,\hat{s}^{i}_L \right),
\end{align}
where $ \hat{s}^{i}_l \in \mathbf{s}_{i} $ and $ \hat{s}^{i}_1 \le \hat{s}^{i}_2 \le \cdots \le \hat{s}^{i}_L $. As shown in Fig.~\ref{fig:Arttribute distributions } (b), both the mean and standard deviation increase as the feature index grows after the sort operation. This observation suggests that the sorted features have more different distributions, which will bring more diversity in the random feature selection and thus result in more biased and unrelated trees. Finally, given a similarity vector $ \hat{\mathbf{s}}_{i} $, any tree classifier $ \mathcal{T}_{h} (\cdot) $ outputs a label:
\begin{align}
\mathcal{T}_{h} (\hat{\mathbf{s}}_{i})=\left\lbrace 
\begin{array}{lcl}
1 \: , & \text{predicted as a fake robot,} \:  \\
0 \: , & \text{predicted as a real robot.} \:
\end{array}
\right.
\end{align}

In the second stage, we aggregate $ H $ intermediate results $ \left\lbrace \mathcal{T}_{1} (\hat{\mathbf{s}}_{i}), \mathcal{T}_{2} (\hat{\mathbf{s}}_{i}), \cdots,  \mathcal{T}_{H} (\hat{\mathbf{s}}_{i}) \right\rbrace $ from all decision trees to make our final prediction. For simplicity, we consider that the importance of each tree is equal to $ \frac{1}{H} $, and then select the label with majority votes as the final outcome. Formally, given a similarity vector $ \mathbf{s}_{i} $, our RF model $ \mathcal{RF} (\cdot) $ predicts the legitimacy of the $ i^{\text{th}} $ robot as 
\begin{align} \label{decision rule}
	\mathcal{RF}\left( \mathbf{s}_{i}\right) =\left\lbrace 
	\begin{array}{lcl}
	1 \: , & \text{if} \frac{1}{H} \sum^{H}_{h=1} \mathbb{I}\left( \mathcal{T}_{h} ( \hat{\mathbf{s}}_{i}) = 1 \right)   \ge 0.5, \:  \\
	0 \: , & \text{otherwise.}   \:
	\end{array}
	\right.
\end{align}
In Eq.~\eqref{decision rule}, $ \mathbb{I}(\cdot) $ represents the indicator function and $ \hat{\mathbf{s}}_{i} $ is the sorted version of $ \mathbf{s}_{i} $.

\section{Implementation and Evaluation} \label{sec:experiment}

\begin{figure}
	\centering
	\includegraphics[width=0.99\linewidth]{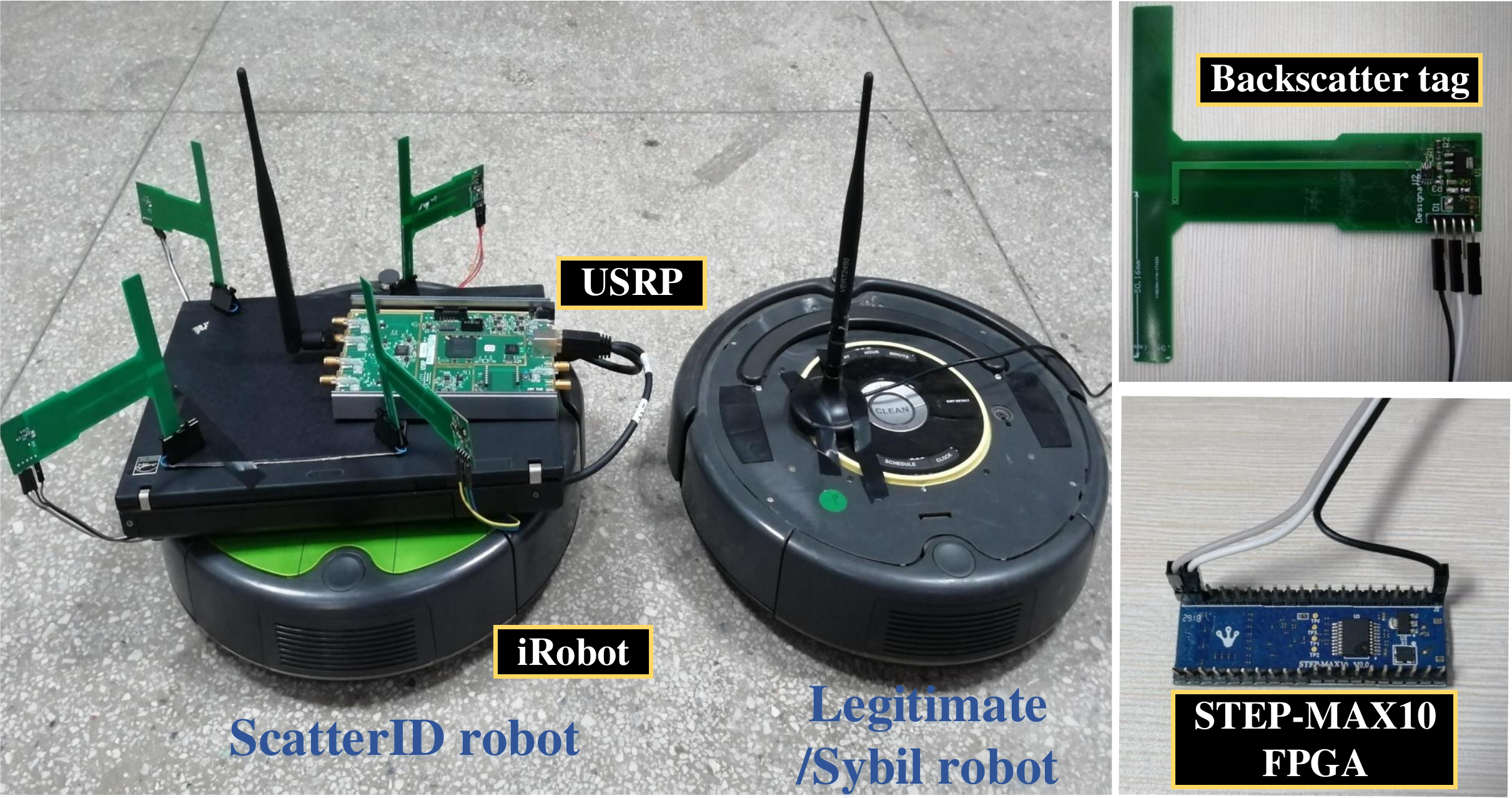}
	\caption{Experimental platform. We implement our system using USRP nodes, backscatter tags and iRobot Create robots. All tags are controlled by an Altera STEP-MAX10 FPGA.}
	\label{fig:platform}
\end{figure}

\subsection{Implementation}
As shown in Fig.~\ref{fig:platform}, we implement our system using iRobot Create robots, backscatter tags, and GNURadio/USRP B210 nodes. We prototype four backscatter tags using off-the-shelf circuit components according to~\cite{liu2013ambient}. In particular, each tag is equipped with one omnidirectional antenna with 3 dBi gain, and it uses the ADG902 RF switch to alternate its antenna impedance between two states for absorbing and reflecting incident signals. All tags are controlled by an Altera STEP-MAX10 FPGA to backscatter signals with a frequency shift of 20~MHz for avoiding interference between the carrier and backscattered signals and transmit data with a bitrate of 4 Kbps in succession for avoiding overlapping among backscattered signals from different tags. Moreover, a total of five commercial iRobot Create robots are leveraged throughout the experiment. Specifically, to build a ScatterID robot, one robot is equipped with four backscatter tags and a USRP node that has a single antenna. The USRP node is surrounded by tags at a distance of 12~cm, and all tags are placed about 15~cm away from each other, which can avoid similar propagation signatures of all tags when backscattering surrounding signals~\cite{tse2005fundamentals}. The left four robots, each of which is equipped with one USRP node that has one antenna, act as neighboring robots of the ScatterID robot. Among neighboring robots, two of them are used as Sybil attackers and two as legitimate robots. In addition, all USRP nodes carried by robotic platforms are configured to communicate in the 2.4~GHz ISM band.

\subsection{Evaluation Methodology}
\begin{figure}
	\centering
	\includegraphics[width=0.99\linewidth]{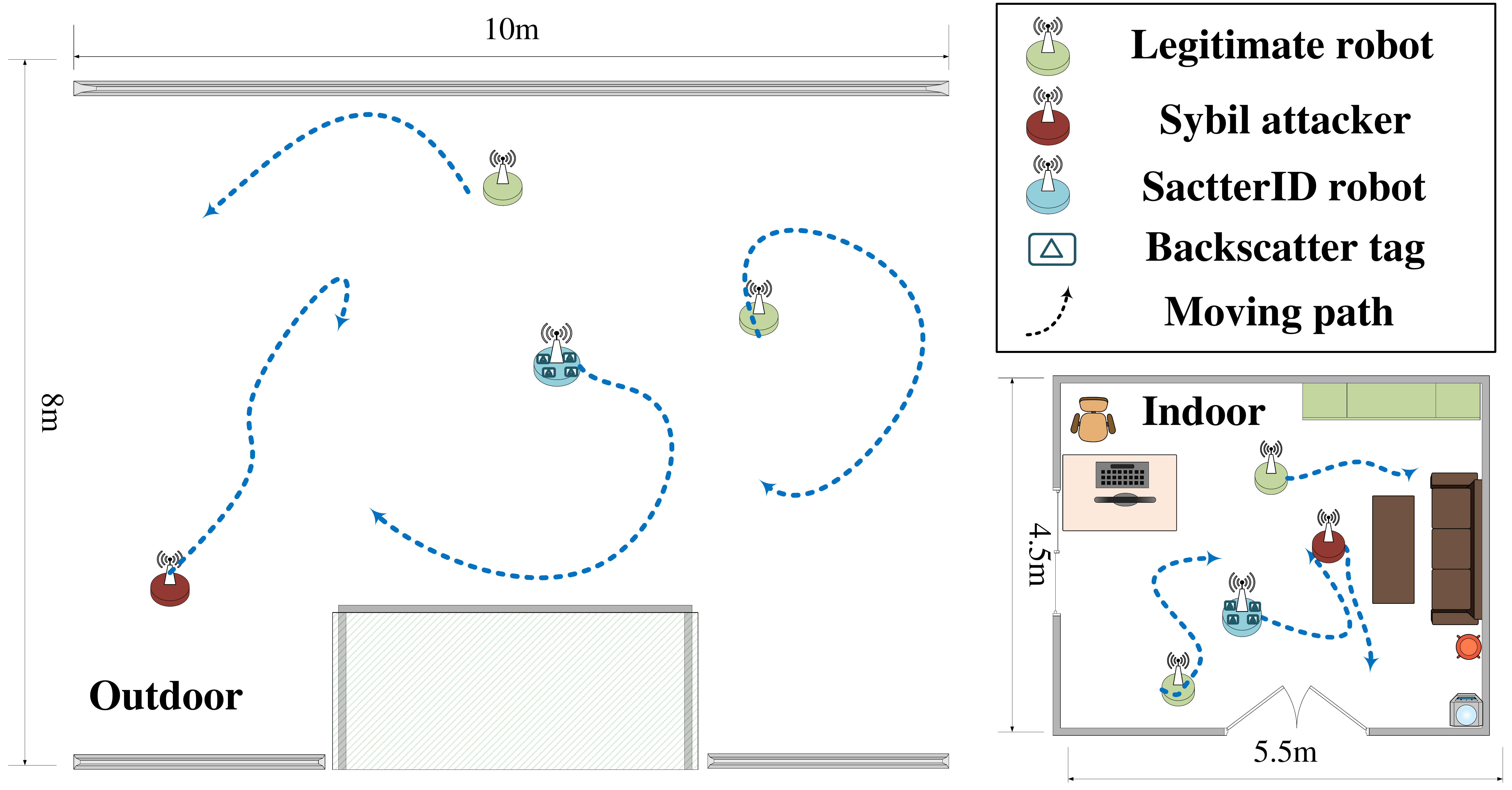}
	\caption{Floor plans of experimental environments. We evaluate our system in the indoor and outdoor environments.}
	\label{fig:environmentsetting}
\end{figure}

\textbf{Experimental Setup.} In our experiment, we evaluate the performance of ScatterID in both indoor and outdoor environments. Specifically, the experimental settings include an office room with a size of 4.5~m $ \times $ 5.5~m and a building rooftop with an area of 8~m $ \times $ 10~m as depicted in Fig.~\ref{fig:environmentsetting}. The office room is a typical multipath-rich environment, which contains some furniture and surrounding walls. The rooftop is a multipath-poor area with a few walls and roof guardrails. According to the speed configurations in~\cite{Snape2009independent,Cheng2018development}, we set all robots to move with a speed of 20~cm/s for easily controlling them at the same time. Based on this moving speed, we set the ScatterID robot to update multipath signatures in every 0.6~s, which can avoid similar signal propagation of successive signatures and keep their distinguishability~\cite{tse2005fundamentals}. Note that generally, the shorter the updating interval, the higher the detection accuracy. However, based on the above analysis, the interval that is shorter than 0.6~s will increase the information redundancy in a signal profile and consequently lead to a marginal accuracy gain. To launch basic Sybil attacks, a robotic attacker broadcasts signals with two or three IDs at the same time. Moreover, it can trigger power-scaling attacks by changing its transmission power for each fake ID using a random scaling coefficient from the set $ \left\lbrace 0.3, 0.6, 0.9 \right\rbrace  $. To launch colluding attacks, we first set two Sybil attackers to together use a total of four fake IDs. Then, in each transmission, one attacker is controlled to randomly take two of these fake IDs to communicate with the ScatterID robot, and the other attacker is set to use the left two fake IDs for communication. Finally, we conduct our experiment over different days in the two environments and yield backscattered signal traces of six-hour in total.

\textbf{Datasets.} We extract signal profiles from the collected radio traces and compute their similarity vectors as described in Section~\ref{sec:system design}. Then, we label the similarity vectors based on corresponding robot legitimacy and obtain about 20K labeled samples. Therein, about 60\% of them are positive and about 40\% are negative. For better evaluating our RF model, we construct two datasets -- Dataset A and Dataset B. Dataset A contains about 15K labeled samples that are collected under basic and power-scaling attacks. Dataset B contains about 5K labeled samples that are collected under colluding attacks.

In our experiment, we implement our RF model on Matlab. During evaluation, we use dataset A for training and testing. Specifically, we randomly partition dataset A into three subsets and use 3-fold cross-validation for evaluating our RF model. In each subset, samples from indoor and outdoor environments are included. Moreover, we leverage dataset B for testing only to verify the system's resistance to colluding attacks. 

\textbf{Evaluation Metrics.} To demonstrate the performance of the proposed system, we use the following metrics.
\begin{itemize}
	\item \textbf{Accuracy.} It is computed as the ratio of the total number of robots that are correctly recognized to the number of legitimate and fake robots.
	\item \textbf{True positive rate (TPR).} It is the ratio of the number of fake robots that are successfully detected to the total number of fake robots. 
	\item \textbf{False positive rate (FPR).} It is the ratio of the number of legitimate robots that are mistakenly recognized to the total number of legitimate robots.
	\item \textbf{Receiver operating characteristic (ROC) curve.} It is a curve in terms of TPRs and FPRs with varying discrimination thresholds in $ \left[0,1 \right]  $. A ROC curve with high TPRs and low FPRs stands for a good binary classifier.
	\item \textbf{Area under the ROC curve (AUROC).} It is a numeric metric to measure the discrimination performance of a binary classifier and computed as the area under ROC curve, falling into $ \left[0.5,1 \right]  $. The closer the AUROC is to one, the better performance is achieved by a classifier.
\end{itemize}
As 3-fold cross-validation is used during evaluation, we average all metric values in 3-fold validations as the final results.

\begin{figure}
	\centering
	\includegraphics[width=0.99\linewidth]{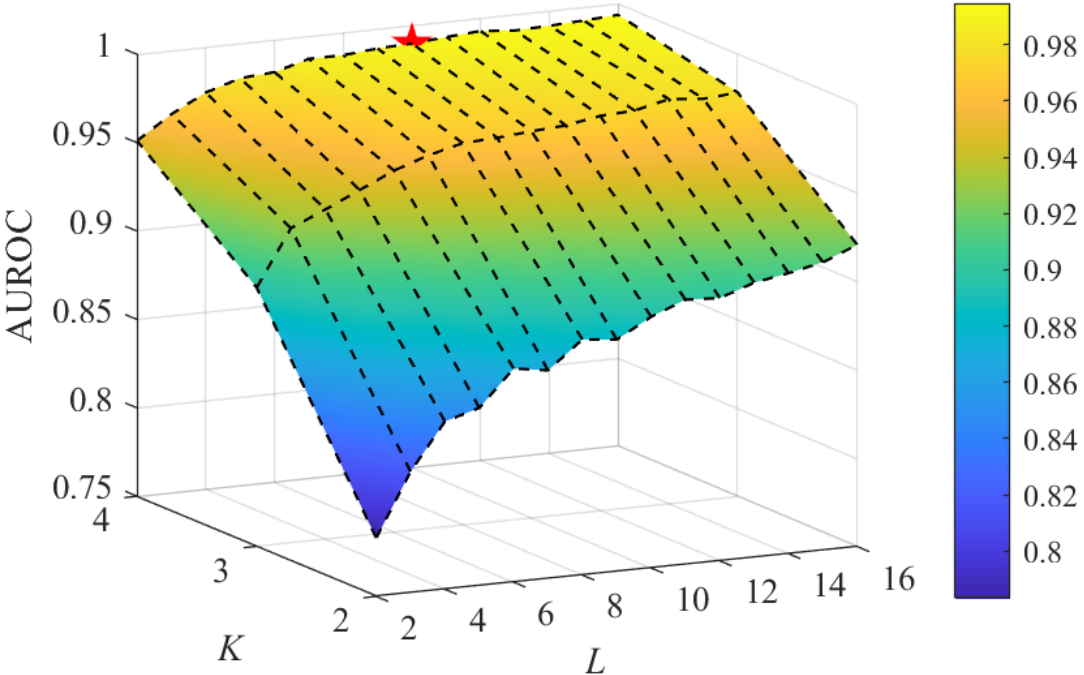}
	\caption{AUROCs in terms of different profile sizes. The red mark corresponds to the selected size with $ K=4 $ and $ L=10 $.}\label{fig:tagnumsfig}
\end{figure}

\subsection{Experimental Results}

\textbf{Profile Size Determination.} Since a signal profile $ \mathbf{F} $ is a fundamental unit for similarity measurement and identity verification, the first step of our experiment is to decide its size. According to Eq.~\eqref{eq: definition of profile}, the profile size is dependent on two parameters $ K $ and $ L $, i.e., the numbers of backscatter tags and multipath signatures, respectively. Generally, the larger the profile size is, the more information will be provided and thus the better performance will be achieved by our system. However, increasing the profile size will also incur more computation complexity and time consumption. Thus, a small-size profile that guarantees high performance is desired. Considering that Sybil attacker detection is a binary classification task, we leverage the AUROC metric to comprehensively measure the system performance under different profile sizes for effective profile size determination. For this purpose, we first extract signal profiles with different sizes from the collected radio traces by varying $ K $ and $ L $ from 2 to 4 and 16, respectively. Then, we train and test our RF model on corresponding profile samples. Next, we calculate ROC curves in terms of different size parameters and compute corresponding AUROC values. As depicted in Fig.~\ref{fig:tagnumsfig}, we observe that AUROC has the lowest value 0.783 with $ K=2 $ and $ L=2 $ and the highest value 0.994 with $ K=4 $ and $ L=16 $. In addition, AUROC value basically rises up as both $ K $ and $ L $ increase, which is in line with the above analysis. Moreover, for the number of tags, four tags ensure that all AUROC values are more than 0.95. For the number of signatures, AUROC grows quickly at the beginning and has marginal growth when $ L $ exceeds 10. Therefore, we choose ten successive multipath signatures from four backscatter tags as a signal profile.

\begin{figure}
	\centering
	\includegraphics[width=0.7\linewidth]{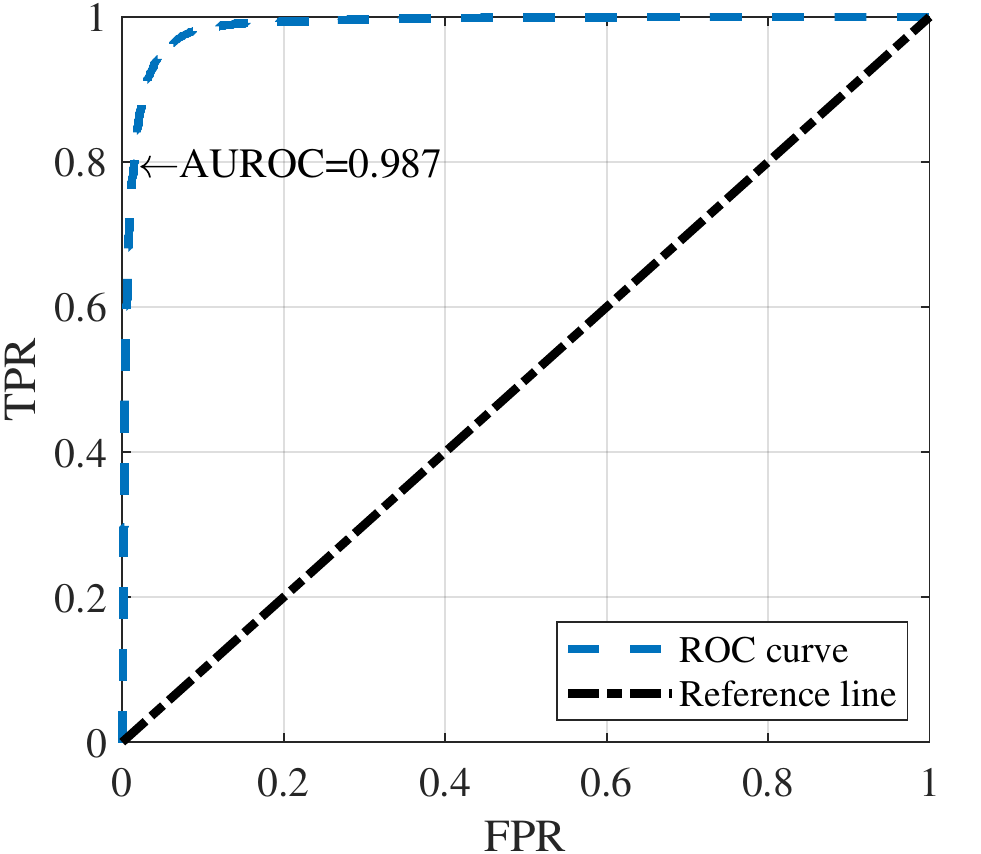}
	\caption{ROC curve of our system. Its AUROC reaches to 0.987 and the reference line represents the random guessing model. }\label{fig:roccurve}
\end{figure}

\begin{figure}
	\centering
	\includegraphics[width=0.99\linewidth]{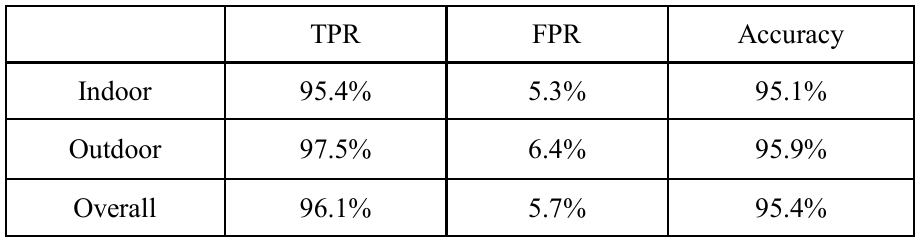}
	\caption{System performance in different environments.}\label{fig:overallperformance}
\end{figure}

\textbf{Overall Performance.} Based on the determined profile size, we illustrate the overall performance of ScatterID in our experiment. First, we show the ROC curve of our system as well as its AUROC to present its capability of discriminating legitimate robots and Sybil attackers. As depicted in Fig.~\ref{fig:roccurve}, the ROC curve closely follows the left-hand and top borders of the ROC space, which suggests that high TPRs and low FPRs are mainly achieved when the discrimination threshold varies in $ \left[0,1 \right]  $. Accordingly, the ROC curve yields a high AUROC of 0.987, which is close to the ideal case. To further illustrate the effectiveness of our system, we summarize its performance in the indoor and outdoor environments as well as the overall performance with respect to TPR, FPR and accuracy in Fig.~\ref{fig:overallperformance}. Compared with the outdoor environment, the indoor setting has much richer multipath due to scattering and reflections from surrounding furniture items and walls. As shown in Fig.~\ref{fig:overallperformance}, such multipath effects make our system obtain a lower TPR in the indoor setting. This is due to the fact that rich multipath propagations disturb backscatter signatures from the same attacker and render them less similar to each other, which consequently causes more positive samples to be mistakenly recognized as negative ones. In the same way, the multipath effects enlarge the discrepancy of negative samples and thus lead to a lower FPR. Furthermore, environment noise in backscattered signals will impact ScatterID's performance. In general, the lower the signal-to-noise ratio (SNR) of backscattered signals, the less similarity between multipath signatures of two fake robots is. To address this issue, our system adopts a moving average method to remove signal outliers, subtracts reflected samples from non-reflected ones to eliminate environmental reflections, and takes a signature sequence as a detection unit to enhance the similarity between two illegitimate signatures under environment dynamics. As Fig.~13 shows, ScatterID achieves high performance in each environment with an accuracy more than 95\%, which demonstrates its robustness to environment noise. To sum up, our system has a detection accuracy of 95.4\%. In particular, it can successfully mitigate 96.1\% of Sybil attacks and correctly recognize 94.3\% of legitimate traffic. The above results show the effectiveness and robustness of the proposed system in discriminating between legitimate and fake robots.

\begin{figure}
	\centering
	\includegraphics[width=0.99\linewidth]{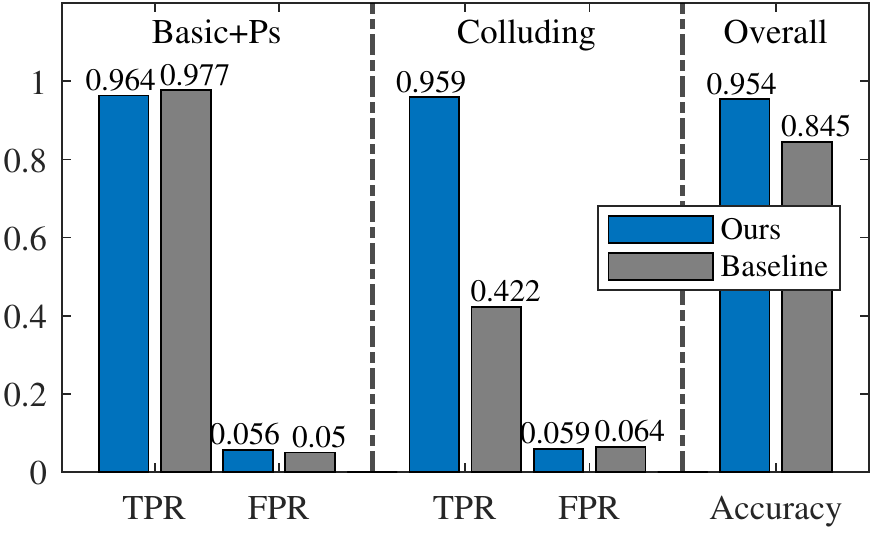}
	\caption{Performance of our and baseline models under various Sybil attacks. Ps stands for power-scaling attacks.}
	\label{fig:oursbaselineperformancecomparison}
\end{figure}

\textbf{Performance under Various Sybil Attacks.} Next, we show that our system has resistance to various Sybil attacks. To illustrate these merits, we set up a baseline system that is developed in our previous work~\cite{yong2020sybil}. In the baseline system, attackers with basic and power-scaling ability are taken into consideration. Similarly, we train and test the baseline model on the backscatter traces under basic and power-scaling attacks with 3-fold cross-validation, and use the traces under colluding attacks for testing only.

Fig.~\ref{fig:oursbaselineperformancecomparison} illustrates the detection performance of our and baseline systems under various types of Sybil attacks. As the figure shows, we can observe that our system has very close performance in the two attack scenarios and achieves a TPR of more than 95\% and a FPR of less than 6\% in each scenario. This is due to that our system extracts representative features from a distance matrix to form a similarity vector, which is resilient to ID switching between colluding attackers as explained in Section~III. However, colluding attacks still render a TPR decrease by 0.5\% and a FPR increase by 0.3\%. These small gaps may be caused by the fact that samples under colluding attacks are never used in the training phase, making our system obtain slightly low performance in the colluding scenario. In addition, when comparing our system with the baseline, we have observed that although our system has a slight performance degradation with a TPR decline of 1.3\% and a FPR increase of 0.6\% under basic and power-scaling Sybil attacks, it achieves a significant performance improvement with a TPR increase of 53.7\% and a FPR decrease of 0.5\% under colluding Sybil attacks. Moreover, when it comes to the overall performance, our system obtains an accuracy increase of 10.9\% in comparison with the baseline. The reason is that in colluding attacks, a Sybil attacker can collude with others and use a random subset of fake IDs in each transmission, which consequently makes itself hard to be detected by the baseline system. Furthermore, the two systems obtain low FPRs in two attack scenarios, because the attacks launched by Sybil attackers have little impact on the system's capability of correctly recognizing legitimate robots. Based on the above observations, we conclude that our system is resilient to various Sybil attacks and has an improvement in overall performance when compared with the baseline.

\begin{figure}
	\centering
	\includegraphics[width=0.99\linewidth]{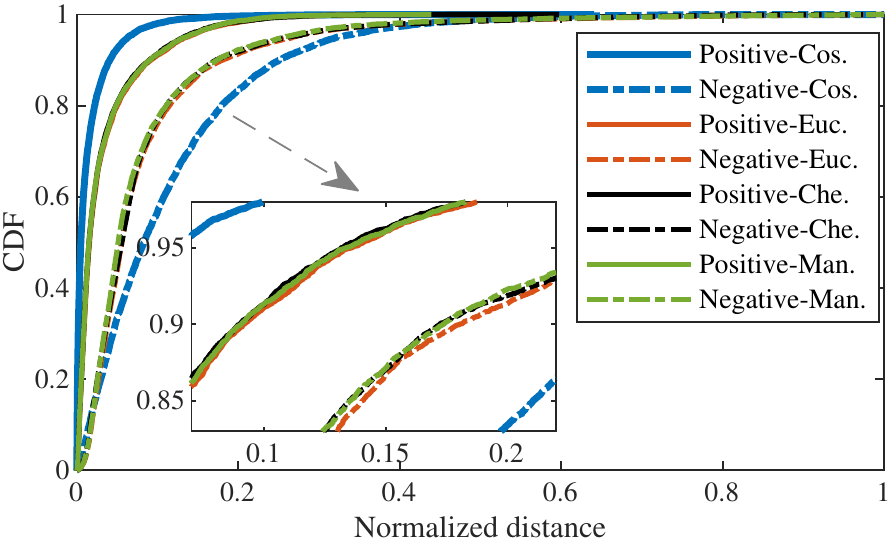}
	\caption{CDFs of positive and negative samples using Cosine, Euclidean, Chebyshev and Manhattan distances.}
	\label{fig:distancecdf}
\end{figure}

\begin{figure}
	\centering
	\includegraphics[width=0.99\linewidth]{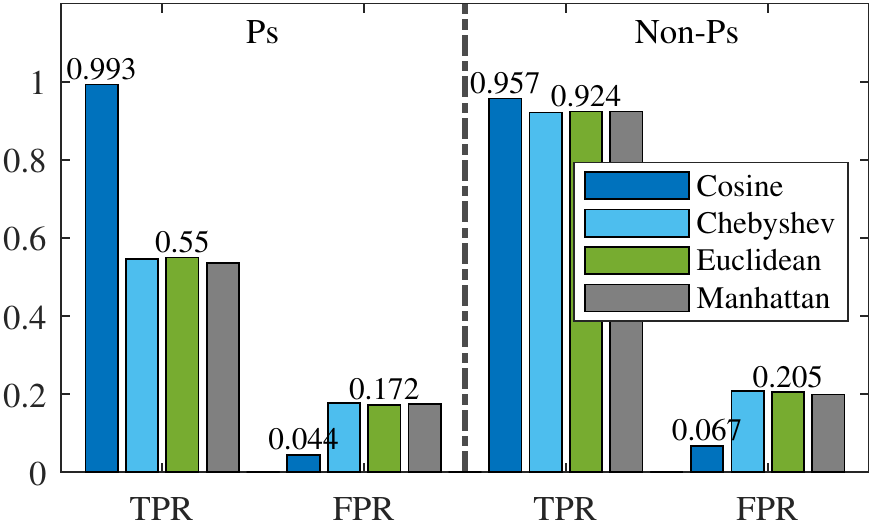}
	\caption{Performance using different distance metrics. Ps stands for power-scaling attacks. We only annotate TPRs and FPRs of Cosine and Euclidean distance metrics for brevity.}
	\label{fig:distanceperformance}
\end{figure}

\textbf{Effectiveness of Cosine Distance.} We further check the effectiveness of Cosine distance used in profile distance measurement. In our application, Cosine distance has an advantage over other distances with its ability to mitigate power-scaling attacks. To illustrate this advantage, we calculate similarity vectors based on Eq.~\eqref{eq: similarity vector} using Cosine, Euclidean, Chebyshev and Manhattan distances, respectively, and obtain four sample datasets corresponding to different distance metrics. Then, we normalize all feature points in the four datasets and compute the cumulative distribution functions (CDFs) of their positive and negative samples, respectively.

As plotted in Fig.~\ref{fig:distancecdf}, the positive CDFs of four metrics are basically higher than the negative ones, because positive samples are from Sybil attackers and their distance features are generally smaller. However, it can be clearly observed that Cosine distance has a steeper positive CDF and a flatter negative CDF in comparison to the other three distances. The observation implies that Cosine distance that captures angular differences is more effective in measuring distances between any two backscatter signatures under power-scaling attacks. To further demonstrate this, we train and test our RF model on the Cosine, Euclidean, Chebyshev and Manhattan datasets, respectively, and report its performance with and without power-scaling attacks in Fig.~\ref{fig:distanceperformance}. As the figure shows, Cosine distance achieves comparable performance in the two attack scenarios and has a TPR larger than 95\% and a FPR smaller than 7\% in each scenario. This is because that in each transmission, backscattered signal strengths are proportional to transmission power, and thus a varying coefficient can be eliminated during Cosine distance calculation. However, it can be found that Euclidean distance obtains a very low TPR of 55\% in the power-scaling scenario, which indicates that it can not handle such attacks. In addition, the FPRs of Euclidean distance in the two attack scenarios are relatively high, which renders it inapplicable in reliably recognizing legitimate robots. The same observations can be found in Chebyshev and Manhattan distances. To conclude, Cosine distance is an effective metric in profile distance measurement and makes our system resilient to power-scaling attacks.

\begin{figure}
	\centering
	\includegraphics[width=0.99\linewidth]{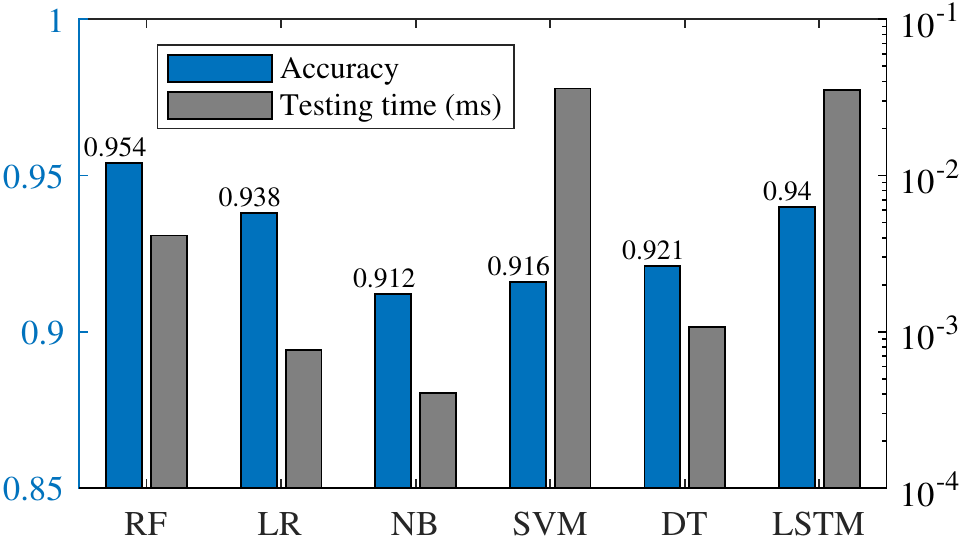}
	\caption{Overall detection accuracy and testing time for each sample using different machine learning models.}
	\label{fig:mlperformance}
\end{figure}

\textbf{Effectiveness of Our RF Model.} Finally, we demonstrate the effectiveness of our RF model for fake robot detection. We compare our RF model with other machine learning models. To do this, we choose several widely-used models, i.e., logistic regression (LR), naive Bayes (NB), support vector machine (SVM), decision tree (DT) and long-short-term memory (LSTM) network with 30 hidden units, as baselines for comparison. Similarly, we use two-thirds of Dataset A for training baselines and the left third as well as Dataset B for testing them. We report their overall detection accuracy as well as run time for each testing sample in Fig.~\ref{fig:mlperformance}. It can be observed that although NB achieves the lowest testing time less than 0.001ms per sample, it also has the lowest accuracy of 91.2\%. Moreover, SVM obtains the highest testing time of about 0.036ms per sample, while having a low detection accuracy of 91.6\%. However, our RF model achieves the highest accuracy of 95.4\% and has a moderate testing time of about 0.004ms per sample. Note that with more hidden units, LSTM network is expected to achieve a better detection accuracy, while inevitably incurring higher computational overhead. Considering the above observations, we conclude that the RF model achieves the best performance compared with the other five machine learning models in our application. 

\begin{figure}
	\centering
	\includegraphics[width=0.99\linewidth]{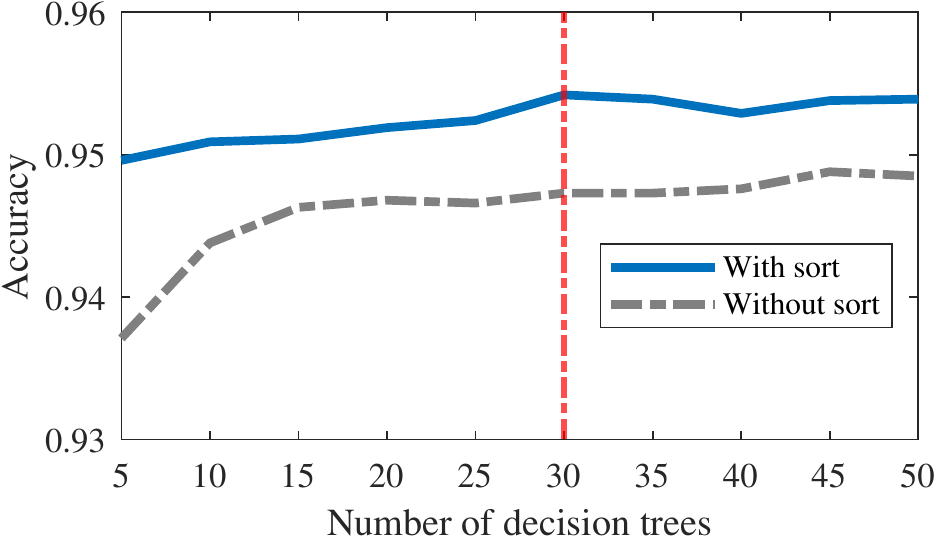}
	\caption{Impact of the sort operation and number of decision trees. The dashed vertical line represents the number of trees used in our RF model.}
	\label{fig:rfperformance}
\end{figure}

Moreover, we present the impact of the sort operation and the number of decision trees on our RF model. For this purpose, we train and test our model with and without the sort operation multiple times by changing the number of decision trees from 5 to 50, and report the accuracy on Dataset B in Fig.~\ref{fig:rfperformance}. Note that the sort operation is introduced in our RF model for differentiating feature distributions and boosting the effectiveness of random feature selection as mentioned in Section~\ref{sec:system design}. As the figure shows, the system with the sort operation always has a higher accuracy on Dataset B. Considering that Dataset B is never used for training, we conclude that the sort operation is capable of improving the generalization performance of our system. Additionally, the number of decision trees in a RF model has an impact on its prediction ability. Generally, the more decision trees are adopted, the better performance is achieved. As shown in Fig.~\ref{fig:rfperformance}, we observe that the system's accuracy rises up as the tree number increases at first, and then it becomes relatively steady after 30 trees. Based on this observation, we adopt 30 decision trees in our RF model.  

\section{Related Work} \label{sec:related work}
\textbf{Multi-Robot Networks.} Recent years have witnessed the proliferation of multi-robot networks in many emerging applications such as surveillance~\cite{rybski2002performance}, exploration~\cite{corah2019distributed} and coverage~\cite{gil2015adaptive,ghaffarkhah2014dynamic}. In these applications, one of the key research issues is how to effectively cooperate among robots to achieve high-quality overall performance. However, effective cooperation is based on the assumption that all claimed identities are accurate and trustful, which can be easily undermined by Sybil attackers who can masquerade a myriad of fictitious robots.

\textbf{Backscatter Communications.} Nowadays, backscatters have become one of the most energy-efficient communication primitives. In~\cite{liu2013ambient}, ambient backscatter is originally proposed to facilitate two simple and batteryless tags to communicate with each other. After that, a variety of novel techniques, such as WiFi backscatter~\cite{kellogg2015wi-fi}, FS backscatter~\cite{zhang2016enabling}, FM backscatter~\cite{wang2017fm} and LoRa backscatter~\cite{talla2017lora}, are successively developed to achieve high energy efficiency in backscatter communications. Beyond their benign uses, backscatter has increasingly been used in other applications such as on-body device authentication~\cite{luo2018authenticating} and object tracking~\cite{luo20193d}. In~\cite{luo2018}, backscatter tags are used to shield static Internet of Things (IoT) networks from spoofing attacks, where attackers try to masquerade other legitimate nodes. Differing from~\cite{luo2018}, we consider Sybil attacks in dynamic robotic networks and provide effective resilience to basic and advanced Sybil attacks.

\textbf{Sybil Attack Mitigation.} Sybil attacks have been widely considered in multi-node networks~\cite{newsome2004sybil,wang2018ghost,faria2006detecting}. The past solutions for Sybil attack mitigation are mainly falling into two categories. (i) Cryptographic-based approaches~\cite{ramkumar2005an,Halford2016Energy} assume prior trust among network nodes and require computationally expensive PSK management schemes. These requirements, however, cannot be satisfied in ad hoc and miniaturized robotic platforms. (ii) Non-cryptographic approaches~\cite{demirbas2006an,xiao2009channel-based,liu2015the,faria2006detecting,xiong2013securearray,gil2017guaranteeing} use wireless PHY information to discriminate between real and fake nodes. However, these approaches passively observe PHY features using bulky multiple antennas and thus do not suit to miniaturized robots with limited payload and hardware capabilities. To transcend these limitations, our work adopts backscatter tags to actively capture fine-grained PHY features and constructs sensitive signal profiles that are easily obtainable to single-antenna robots for Sybil attack detection.

\textbf{Machine Learning in Wireless Security.} Machine learning algorithms have been increasingly applied to enhance the security performance of wireless communication systems. In~\cite{shi2017smart}, a neural network is designed to perform user authentication relying on WiFi signals. In~\cite{yong2020authenticating}, an adversarial model is proposed to authenticate on-body devices under various user body motions. Moreover, the work~\cite{xiao2018phy} proposes a logistic regression model to reduce communication overhead between MIMO landmarks and its security agent. Our work develops a RF model that is suitable for backscatter signatures and can reliably detect Sybil attacks in robotic networks.

\section{Conclusion}\label{sec:conclusion}
This paper presents ScatterID, a lightweight system against Sybil attacks for networked and miniaturized robots in many cooperative tasks. Instead of passively measuring PHY features using bulky multiple antennas, ScatterID utilizes featherlight backscatter tags to actively manipulate multipath propagation and creates salient multipath features obtainable to single-antenna robots. These features are used to identify the spatial uniqueness of each moving robot under Sybil attackers with power-scaling and colluding abilities. We implement our system on commercial off-the-shelf robotic platforms and extensively evaluate it in typical indoor and outdoor environments. The experimental results show that ScatterID achieves a high AUROC of 0.987 and an overall accuracy of 95.4\% for identity verification. In addition, it can successfully detect 96.1\% of fake robots while mistakenly rejecting just 5.7\% of legitimate ones.

\bibliographystyle{IEEEtran}
\bibliography{IEEEabrv,./Sybildetection}

\end{document}